\pdfoutput=1
\documentclass[
 pagesize=automedia,fontsize=12pt,
 BCOR=15mm,DIV=22,
 twoside,headinclude,footinclude=false,
 ngerman,fleqn,             
 bibtotocnumbered,          
 liststotoc,               
 listsleft,                
 pointlessnumbers,          
 cleardoublepage=empty, a4paper     
]{article}
\usepackage{ifluatex}
\ifluatex 
\usepackage{polyglossia} 
\setdefaultlanguage{english}

\else 
\usepackage[utf8]{inputenc} 
\usepackage[T1]{fontenc} 
\usepackage[main=english, ngerman]{babel}

\fi

\usepackage{cmap}
\usepackage{setspace} 
\usepackage{graphicx,xcolor}
\usepackage{tabularx}
\usepackage{amsmath,amsfonts,amssymb}
\usepackage{flafter,afterpage}
\usepackage[section]{placeins}
\usepackage[margin=8mm,font=small,labelfont=bf,format=plain]{caption}
\usepackage[margin=8mm,font=small,labelfont=bf,format=plain]{subcaption}

\numberwithin{equation}{section}
\numberwithin{figure}{section}
\numberwithin{table}{section}

\usepackage{tikz-feynman}
\usepackage{rotating}
\usepackage{tikz}
\usetikzlibrary{arrows.meta, fit}
\tikzset{%
	>={Latex[width=2mm,length=2mm]},
	base/.style = {rectangle, rounded corners, draw=black,
		minimum width=4cm, minimum height=1cm,
		text centered, font=\sffamily},
	activityStarts/.style = {base, fill=blue!30},
	startstop/.style = {base, fill=red!30},
	activityRuns/.style = {base, fill=green!30},
	process/.style = {base, fill=orange!15},
	decision/.style={base, diamond, aspect=2, text width=5em, fill=blue!30},
	coord/.style={coordinate, on grid, node distance=6mm and 25mm},
}
\usepackage{listings} 
\usepackage{pdflscape}
\usepackage{color}
\usepackage[colorlinks, pdfpagelabels, pdfstartview = FitH, bookmarksopen = true, bookmarksnumbered = true, linkcolor = black, plainpages = false, hypertexnames = false, citecolor = black, urlcolor=black] {hyperref}
\usepackage[backend=biber, 
style=phys, 
pageranges=false, 
language=english, 
biblabel=brackets,
doi=true,
eprint=true, url=true
]{biblatex} 
\DeclareFieldFormat{doi}{\addcomma\space\url{https://doi.org/#1}}
\usepackage{xurl}
\usepackage{picinpar}
\usepackage{booktabs} 
\usepackage{physics} 
\usepackage{microtype}
\usepackage{bbm} 
\usepackage{tcolorbox}
\usepackage{verbatim} 
\usepackage{tablefootnote}
\usepackage{layout}    
\usepackage{geometry}
\usepackage{titling}
\usepackage{xspace}
\usepackage{enumitem}

\newcommand{\Sherpa}{S\protect\scalebox{0.8}{HERPA}\xspace}
\newcommand{\Comix}{C\protect\scalebox{0.8}{OMIX}\xspace}
\newcommand{\Amegic}{A\protect\scalebox{0.8}{MEGIC}\xspace}
\newcommand{\Madgraph}{M\protect\scalebox{0.8}{AD}G\protect\scalebox{0.8}{RAPH}\xspace}
\newcommand{\Pythia}{P\protect\scalebox{0.8}{YTHIA}\xspace}
\newcommand{\Herwig}{H\protect\scalebox{0.8}{ERWIG}\xspace}
\newcommand{\Phantom}{P\protect\scalebox{0.8}{HANTOM}\xspace}
\newcommand{\Recola}{R\protect\scalebox{0.8}{ECOLA}\xspace}
\newcommand{\MoCaNLO}{M\protect\scalebox{0.8}{O}C\protect\scalebox{0.8}{A}N\protect\scalebox{0.8}{LO}\xspace}

\newcommand{\Collier}{C\protect\scalebox{0.8}{OLLIER}\xspace}
\newcommand{\Rivet}{R\protect\scalebox{0.8}{IVET}\xspace}

\newcommand{\SMCatNLO}{S--M\protect\scalebox{0.8}{C}@N\protect\scalebox{0.8}{LO}\xspace}
\newcommand{\MCatNLO}{M\protect\scalebox{0.8}{C}@N\protect\scalebox{0.8}{LO}\xspace}

\newcommand{\EWvirt}{\ensuremath{\text{EW}_{\text{virt}}}}

\pretitle{\begin{center}\Huge\bfseries} 
\posttitle{\end{center}\vspace*{5mm}} 
\title{Polarised cross sections for vector boson production with \Sherpa}
\author{
	Mareen Hoppe\textsuperscript{1},
	Marek Schönherr\textsuperscript{2},
	Frank Siegert\textsuperscript{1}\\ 
	\textsuperscript{1}{\small{Institute for Nuclear and Particle Physics, TUD Dresden University of Technology, D–01062 Dresden, Germany}}\\ 
	\textsuperscript{2}{\small{Institute for Particle Physics Phenomenology, Department of Physics, Durham University, Durham, DH1 3LE, UK}}\\ 
}
\date{} 
\renewcommand{\maketitlehookd}{%
	\begin{abstract}
		\noindent Measurements of vector boson polarisation in vector boson production processes offer a powerful probe of the electroweak symmetry breaking mechanism, scrutinising the Standard Model and new physics scenarios alike. Since massive vector bosons can only be observed as intermediate particles, polarised cross section templates from simulation are necessary to extract their polarisation from measurable unpolarised distributions. In this work we present an extension of the \Sherpa Monte-Carlo event generator allowing the simulation of polarised cross sections for vector boson production processes. Based on the narrow-width approximation, polarised cross sections of all possible polarisation combinations for an arbitrary number of intermediate vector bosons can be simulated in a single simulation run. In addition, it is possible to directly predict the interference between different intermediate polarisation states, and various differing polarisation definitions can be studied simultaneously. Besides the simulation of polarised cross sections at fixed LO and LO+PS accuracy as well as in multijet merged calculations, we also present parton-shower-matched polarised cross sections with approximate NLO QCD corrections in the vector boson production processes. We demonstrate that the differences of this approximation to full NLO QCD predictions are small and it thus opens up the possibility for fully-simulated calculations at the hadron level including polarisation information and higher-order QCD effects for the first time.
	\end{abstract}
}

\begin{document}
	\maketitle
	\thispagestyle{empty}
	\unitlength = 1mm
	\begin{picture}(0,0)
	  \put(120,160){IPPP/23/60, MCNET-23-11}
	\end{picture}

	\newpage
	\tableofcontents

	\section{Introduction}
\label{chapter: Introduction}
The investigation of the polarisation of massive vector bosons (VB) has gained significant attention in recent years, both theoretically and experimentally. The longitudinal polarisation of massive VBs is a direct consequence of the electroweak symmetry breaking mechanism, making polarised VB production a very promising group of processes for probing this mechanism. Similarly, the diagrams of many VB production processes contain triple and quartic gauge coupling vertices, which further contribute to their significance as probes for the innermost gauge symmetry structure of the Standard Model (SM).
In addition, measurements of VB polarisation have the potential to provide insights into physics beyond the Standard Model (BSM). In such models, for example, modifications in the VB scattering (VBS) cross sections of longitudinally polarised $\text{W}^\pm$ and Z bosons can arise due to different Higgs boson couplings to gauge bosons or the presence of new resonances \cite{Espriu2012, Chang2013}. Some new physics models even predict differences in the VBS cross sections of transversely polarised $\text{W}^\pm$ and Z bosons \cite{Brass2018}.

First VB polarisation measurements at the LHC are conducted with data from collisions at 7 and 8 TeV center-of-mass energy (CME) for  $\text{W}^\pm$ boson+jet and Z boson+jet production \cite{CMSCollaboration2011, ATLASCollaboration2012a, CMSCollaboration2020a, CMSCollaboration2015, ATLASCollaboration2016a} as well as in top quark decays~\cite{ATLASCollaboration2016, CMSCollaboration2016, CMSATLASC2020}. Data taken from the recently finished Run 2 of the LHC at 13 TeV CME is currently being analysed. First measurements are presented for $\text{W}^\pm$Z production \cite{ATLASCollaboration2019, CMSCollaboration2021, ATLASCollaboration2022} and $\text{W}^\pm \text{W}^\pm$ scattering \cite{CMSCollaboration2020}. The expected high luminosity in the forthcoming LHC-runs will provide higher sensitivity to VB polarisation and will also enable polarisation measurements of very rare processes such as the various VBS modes \cite{Azzi2019, CMS-PAS-FTR-18-014}.

Since massive VBs only appear as intermediate particles in observable processes, VB polarisation measurements require polarised cross section templates provided by Monte Carlo (MC) event generators to extract polarisation information from the unpolarised, measurable data.
Currently, only a few generators are able to separate polarisation states on amplitude level: The \Madgraph \cite{Alwall2014, Franzosi2019} matrix element generator is able to simulate polarised cross sections for general multi-boson processes at leading order~(LO) and interface these to parton shower programs like \Pythia~\cite{Bierlich2022} and \Herwig~\cite{Bahr2008, Bellm2016}. The MC event generator \Phantom \cite{Ballestrero2009} can provide LO polarised predictions for $2\rightarrow6$ processes. With \Recola \cite{Actis2012, Actis2016} and \Collier \cite{Denner2016}, the generation of polarised events has been extended to next-to-leading order~(NLO) QCD for diboson-production processes in fully- and semi-leptonic decay channels \cite{Denner2020, Denner2021, Denner2021a, Denner2022} and to NLO EW in Z boson pair production \cite{Denner2021a}. For $\text{W}^\pm$ boson+jet- and $\text{W}^+\text{W}^-$ boson production predictions up to NNLO QCD \cite{Pellen2022, Poncelet2021}, for inclusive $\text{W}^\pm$Z boson pair production up to NLO QCD+EW \cite{Le2022, DNLJB2022, Dao2023} are available. \\
The framework introduced in this work enables the simulation of polarised cross sections for unstable VBs with the general-purpose MC event generator \Sherpa \cite{Bothmann2019}. It thus provides a second fully realistic prediction at the hadron level including effects of parton showers and hadronisation. Polarised cross sections of all possible polarisation combinations can be computed in one simulation run and are provided as additional event weights in \Sherpa. Furthermore, the interference between different polarisations can be calculated directly without relying on histogram subtraction methods, and several polarisation definitions are provided.
The implementation relies on tree-level matrix elements for multi-leg matrix elements, and it is shown that these can be utilised to simulate the majority of the effect on VB polarisations at NLO QCD with \Sherpa if the influence of virtual corrections on polarisation fractions is negligible.

This paper is organised as follows. In Sec.~2, the definition of polarisation for intermediate VBs is introduced. Implementation details of the new polarisation framework in \Sherpa are presented in Sec.~3, also covering how the simulation of VB polarisation aspects at NLO QCD becomes possible. The implementation is validated against literature data at fixed LO for several processes in Sec.~4. First applications of the new framework in phenomenological analyses investigating higher-order QCD corrections to polarised cross sections are discussed in Sec.~5. Finally, Sec.~6 gives a summary of this work and an outlook into future extensions of the new framework and planned applications.

	\section{Definition of polarised amplitudes for intermediate vector bosons} \label{Sec.:VB polarisation definition}

In this section we introduce the definition of the polarisation of intermediate massive VBs.
The production of a single massive VB and its subsequent decay into a fermion pair is described in unitary gauge by the amplitude,
\begin{equation} 		
	\mathcal{M}=\mathcal{M}_\mu^{\mathrm{prod}} \Bigg(\frac{i(-g^{\mu \nu}+\frac{q^\mu q^\nu}{m_V^2})}{q^2-m_V^2+i\Gamma_V m_V}\Bigg) \mathcal{M}_\nu^{\mathrm{decay}}\,,
	\label{Gl.: Matrix element Wj production and decay}
\end{equation}
with $m_{\text{V}}$, $\Gamma_{\text{V}}$ and $q^\mu$ denoting mass, width and four-momentum of the intermediate VB. \\
This amplitude is connected with the VB polarisation, described by four polarisation vectors $\varepsilon^\mu_\lambda(q)$, via the completeness relation
\begin{equation}
	\Biggl(-g^{\mu \nu}+\frac{q^\mu q^\nu}{m_V^2}\Biggr)=\sum_{\lambda=1}^{4} \varepsilon^\mu_\lambda(q) \varepsilon^{\ast \nu}_\lambda (q)\,. 
	\label{Gl.: Completness relation}	
\end{equation} 
This sum contains the three physical polarisation states, two transverse and one
longitudinal, and a fourth unphysical polarisation which only vanishes for on-shell states.\footnote{
  The fourth polarisation does also not contribute if the intermediate VB decays into massless leptons \cite{Ballestrero2017}.}
The three physical polarisation vectors have the properties,
\begin{equation}
	q_\mu \cdot\varepsilon^\mu_\lambda(q) = 0 \hspace{2cm} \varepsilon^\mu_\lambda(q) \cdot \varepsilon^\ast_{\mu,\lambda^\prime}(q) = - \delta_{\lambda\lambda^\prime}.
	\label{Eq.: Polarisation vector properties}
\end{equation}
Their form depends on the chosen spin basis. For an on-shell massive VB with momentum \\$q^\mu=(q^0, |\vec{q}| \cos \phi \sin \theta, |\vec{q}| \sin \phi \sin \theta, |\vec{q}| \cos \theta)$ considered in a helicity basis, they are given by
\begin{equation}
	\begin{aligned}
		\varepsilon_\pm^\mu(q) &= \frac{e^{\pm i\phi}}{\sqrt{2}} (0, -\text{cos}\,\theta \,\text{cos}\,\phi\pm i \text{sin}\,\phi, - \text{cos}\,\theta \,\text{sin}\,\phi\mp i \text{cos}\,\phi, \text{sin}\,\theta) \,, \\
		\varepsilon^\mu_0(q) &= \frac{q^0}{m}\Big(\frac{|\vec{q}|}{q^0}, \text{cos}\,\phi \,\text{sin}\,\theta, \text{sin}\,\phi\,\text{sin}\,\theta, \text{cos}\,\theta\Big) \,.
		\label{Eq.: Polarisation vectors in helicity basis}
	\end{aligned}
\end{equation}
Besides this four-vector representation, the polarisation vectors can also be expressed in terms of Weyl spinors. The polarisation vectors implemented in \Sherpa's built-in matrix-element generator \Comix~\cite{Gleisberg2008} take the form~\cite{Dittmaier1998}
\begin{equation}
	\varepsilon_{+, \dot{A}B}(q) = \frac{\sqrt{2}a_{\dot{A}}b_B}{\langle ab \rangle^\ast}\hspace{1cm}\varepsilon_{-, \dot{A} B}(q) = \frac{\sqrt{2}b_{\dot{A}}a_B}{{ \langle a b \rangle} } \hspace{1cm} \varepsilon_{0, \dot{A}B}(q) = \frac{1}{m_V}\big(b_{\dot{A}}b_B-\alpha a_{\dot{A}}a_B\big)\,.
	\label{Eq.: Polarisation vectors Spinorrepresentation}
\end{equation} 
$a_A$ and $b_A$ are Weyl spinors which corresponds to the light-like four-vectors in the decomposition of the VB momentum $q^\mu=\alpha a^\mu + b^\mu$ with $\alpha = \frac{q^2}{2 a\cdot q}$ necessary to express four-vectors in terms of those spinors. The four-vector $a^\mu$ can generally be chosen arbitrarily, however, for massive particles it takes on a physically meaningful role as it fixes the spin axis $s^\mu$ of the particle \cite{Alnefjord2020}
\begin{equation}
	s^\mu = \frac{1}{m}(q^\mu - 2\alpha a^\mu) \,.
	\label{Eq.: Spin axis}
\end{equation} 
The helicity basis is obtained by setting $a^\mu \propto(1,-\vec{q}/|\vec{q}|)$ which results in a spin vector pointing in the direction of $\vec{q}$
\begin{equation}
	s^\mu_{\text{hel}}=\frac{1}{m}\Big(|\vec{q}|, q^0 \frac{\vec{q}}{|\vec{q}|}\Big) \,.
\end{equation}
Then, polarisation vectors calculated in \Sherpa have the form of Eq.~\eqref{Eq.: Polarisation vectors in helicity basis} after transforming them back to the four-vector representation.
At this point it is paramount to note that the representations of polarisation vectors in Eqs.~\eqref{Eq.: Polarisation vectors in helicity basis} and \eqref{Eq.: Polarisation vectors Spinorrepresentation} are not Lorentz-covariant, $\Lambda^\mu\,_\nu\varepsilon^\nu(q, \lambda) \neq \varepsilon^\mu(\Lambda q, \lambda)$. Consequently, the polarisation of a particle depends on the frame in which its polarisation vectors are calculated.

Out of all possible reference frames (at least) two different frames can now be distinguished to be useful in polarisation measurements -- the laboratory rest frame and the rest frame of the massive VBs. Of course, the polarisation vectors obtained in one frame differ from those that are obtained in the other, and contributions of the individual polarisations to the (invariant) unpolarised cross section of the whole process are thus frame dependent.
This fact is used in experimental analyses \cite{ATLASCollaboration2019, CMSCollaboration2020} to maximise the contribution of the interesting polarisation (usually the longitudinal polarisation).

Since massive VBs only appear as intermediate, off-shell particles in measurable processes, their polarisation can only be deduced from the distributions of their final state decay products. For fully leptonic decays, massless leptons and no applied lepton selection criteria, an analytical equation for the angular distribution of the $\text{W}^\pm$/ Z~boson decay products as a function of the lepton decay angle\footnote{The (lepton) decay angle $\theta^\ast$ is defined as the angle between the charged lepton's momentum in the VB rest frame and the VB's flight direction in the reference frame used for polarisation definition, the VB rest frame is reached by boosting the charged lepton's momentum from the reference frame.} exists which allows for a determination of the polarisation fractions by projecting the angular distribution on Legendre polynomials \cite{Ballestrero2017, Ballestrero2019}.

However, in realistic setups, lepton selection criteria need to be applied. They spoil the factorisation of the angular dependence which is necessary to derive this angular distribution. Hence, to measure polarisation fractions in realistic setups, polarised cross sections need to be simulated.

The fact that VBs are only present as intermediate particles leads to two difficulties in the definition of polarised cross sections which would not arise for external VBs. The first one is a direct consequence of Eq.~\eqref{Gl.: Completness relation}: By inserting Eq.~\eqref{Gl.: Completness relation} into Eq.~\eqref{Gl.: Matrix element Wj production and decay}, the matrix element can be factorised into the production and decay of an on-shell VB
\begin{equation}
	\begin{aligned}
			\mathcal{M} &= \frac{i}{q^2-m^2_\text{V}+i\Gamma_\text{V}m_\text{V}} \sum_\lambda \mathcal{M}^\text{prod}_{\mu} \varepsilon^{\ast\mu}_\lambda \varepsilon^{\nu}_\lambda \mathcal{M}^\text{decay}_{\nu} &= \frac{i}{q^2-m^2_\text{V}+i\Gamma_\text{V}m_\text{V}} \sum_\lambda \mathcal{M}^\mathcal{P}_{\lambda} \mathcal{M}^\mathcal{D}_\lambda =: \sum_\lambda \mathcal{M}^\mathcal{F}_\lambda \,,
	\end{aligned}
\end{equation}
with $\mathcal{M}^\mathcal{F}_\lambda$ being the complete amplitude containing a single VB with definite polarisation $\lambda$, and $\mathcal{M}^\mathcal{P}_{\lambda}$ and $\mathcal{M}^\mathcal{D}_\lambda$ being the respective production and decay amplitudes. Squaring this yields
\begin{equation}
	\underbrace{|\mathcal{M}|^2\vphantom{\sum_{\lambda \neq \lambda'}}}_{\text{coherent sum}} = \underbrace{\sum_{\lambda} |\mathcal{M}_\lambda^\mathcal{F}|^2\vphantom{\sum_{\lambda \neq \lambda'}}}_{\text{polarised contributions}, \, \text{incoherent sum}} + \underbrace{\sum_{\lambda \neq \lambda'} \mathcal{M}_\lambda^\mathcal{F}  \mathcal{M}_{\lambda^\prime}^{\ast \mathcal{F}}}_{\text{interference}}\,,
	\label{Gl.: Interference term}
\end{equation}
where not only transition matrix elements with intermediate VBs of definite polarisation contribute, but also interferences between different VB polarisation states emerge. These additional interference contributions are generally non-negligible, and only vanish in the absence of lepton selection criteria \cite{Ballestrero2017}. Otherwise, they need to be considered as an additional part of the polarisation measurement. Unless stated otherwise, when discussing ``interference'' in the following, this will always refer to this interference contribution between different polarisations. Likewise, ``polarised contributions'' form the incoherent sum of the contributions with definite VB polarisation states in Eq.~\eqref{Gl.: Interference term}.

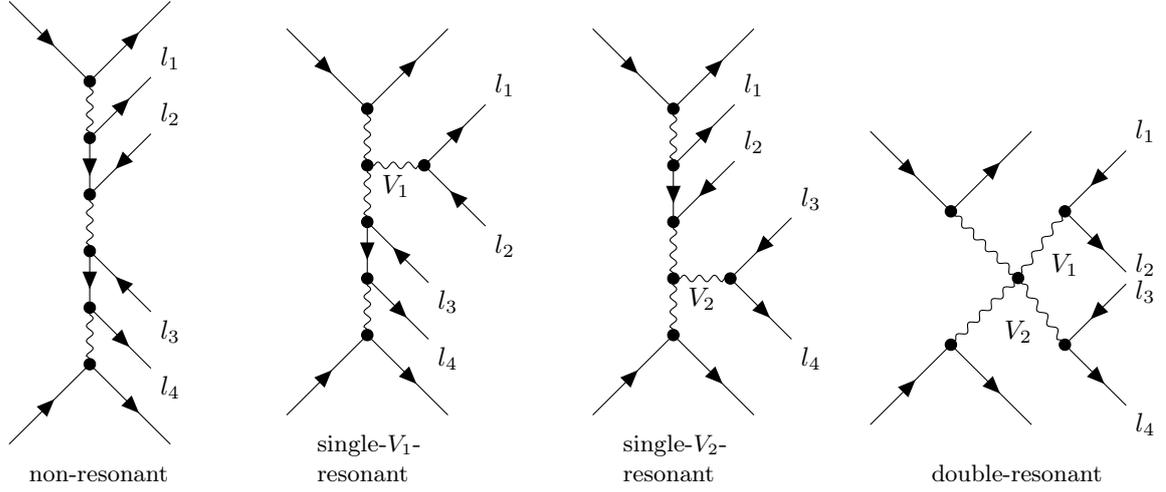
\begin{figure}[h]
	\centering
\begin{subfigure}{0.23\textwidth}
	\centering
 	\begin{tikzpicture}
 		\begin{feynman}
 			\vertex[dot] (a) {};
 			\vertex[above left=of a] (i1);
 			\vertex[below= 0.75cm of a, dot] (b) {};
 			\vertex[below= 0.75cm of b, dot] (c) {};
 			\vertex[below= 0.75cm of c, dot] (d) {};
 			\vertex[below= 0.75cm of d, dot] (e) {};
 			\vertex[below= 0.75cm of e, dot] (f) {};
 			\vertex[below left=of f](i2);
 			\vertex[above right=of a](f1);
 			\vertex[above right=of b](f2){\(l_1\)};
 			\vertex[above right=of c](f3){\(l_2\)};
 			\vertex[below right=of d](f4){\(l_3\)};
 			\vertex[below right=of e](f5){\(l_4\)};
 			\vertex[below right=of f](f6);
 			\diagram*[vertical=a to b]{
 				(i1) -- [fermion] (a) -- [fermion] (f1),
 				(a) -- [boson] (b),
 				(b) -- [fermion] (f2),
 				(b) --[fermion] (c),
 				(f3) -- [fermion] (c),
 				(c) -- [boson] (d),
 				(f4) -- [fermion] (d) --[fermion] (e) -- [fermion] (f5),
 				(e) -- [boson] (f),
 				(i2) -- [fermion] (f) -- [fermion] (f6),
 			};
 		\end{feynman}
 	\end{tikzpicture}
\caption*{non-resonant}
\end{subfigure}
\begin{subfigure}{0.23\textwidth}
	\centering
 	\begin{tikzpicture}
 		\begin{feynman}
 			\vertex[dot] (a) {};
 			\vertex[above left=of a] (i1);
 			\vertex[below= 0.75cm of a, dot] (b) {};
 			\vertex[right= 0.75cm of b, dot] (c) {};
 			\vertex[below= 0.75cm of b, dot] (d) {};
 			\vertex[below= 0.75cm of d, dot] (e) {};
 			\vertex[below= 0.75cm of e, dot] (f) {};
 			\vertex[below left=of f](i2);
 			\vertex[above right=of a](f1) ;
 			\vertex[above right=of c](f2){\(l_1\)};
 			\vertex[below right=of c](f3){\(l_2\)};
 			\vertex[below right=of d](f4){\(l_3\)};
 			\vertex[below right=of e](f5){\(l_4\)};
 			\vertex[below right=of f](f6);
 			\diagram*{
 				(i1) -- [fermion] (a) -- [fermion] (f1),
 				(a) -- [boson] (b),
 				(b) --[boson, edge label'=\(V_1\)] (c),
 				(c) -- [fermion] (f2),
 				(f3) -- [fermion] (c),
 				(b) -- [boson] (d),
 				(f4) -- [fermion] (d) --[fermion] (e) -- [fermion] (f5),
 				(e) -- [boson] (f),
 				(i2) -- [fermion] (f) -- [fermion] (f6),
 			};
 		\end{feynman}
 	\end{tikzpicture}
	\caption*{single-$V_1$-resonant}
\end{subfigure}
\begin{subfigure}{0.23\textwidth}
	\centering
 	\begin{tikzpicture}
 		\begin{feynman}
 			\vertex[dot] (a) {};
 			\vertex[above left=of a] (i1);
 			\vertex[below= 0.75cm of a, dot] (b) {};
 			\vertex[below= 0.75cm of b, dot] (c) {};
 			\vertex[below= 0.75cm of c, dot] (d) {};
 			\vertex[right= 0.75cm of d, dot] (e) {};
 			\vertex[below= 0.75cm of d, dot] (f) {};
 			\vertex[below left=of f](i2);
 			\vertex[above right=of a](f1) ;
 			\vertex[above right=of b](f2){\(l_1\)};
 			\vertex[above right=of c](f3){\(l_2\)};
 			\vertex[above right=of e](f4){\(l_3\)};
 			\vertex[below right=of e](f5){\(l_4\)};
 			\vertex[below right=of f](f6);
 			\diagram*{
 				(i1) -- [fermion] (a) -- [fermion] (f1),
 				(a) -- [boson] (b),
 				(b) --[fermion] (c),
 				(b) -- [fermion] (f2),
 				(f3) -- [fermion] (c),
 				(c) -- [boson] (d),
 				(d) -- [boson, edge label'=\(V_2\)] (e),
 				(f4) -- [fermion] (e),
 				(e) -- [fermion] (f5),
 				(d) -- [boson] (f),
 				(i2) -- [fermion] (f) -- [fermion] (f6),
 			};
 		\end{feynman}
 	\end{tikzpicture}
	\caption*{single-$V_2$-resonant}
\end{subfigure}
\begin{subfigure}{0.23\textwidth}
	\centering
 	\begin{tikzpicture}
 		\begin{feynman}
 			\vertex[dot] (a) {};
      		\vertex[above left=of a] (i1);
 			\vertex[below right= 1.25cm of a, dot] (b) {};
 			\vertex[below left= 1.25cm of b, dot] (c) {};
 			\vertex[right= 1.5cm of a, dot] (d) {};
 			\vertex[right= 1.5cm of c, dot] (e) {};
 			\vertex[below left=of c](i2);
 			\vertex[above right=of a](f1) ;
 			\vertex[above right=of d](f2){\(l_1\)};
 			\vertex[below right=of d](f3){\(l_3\)};
 			\vertex[above right=of e](f4){\(l_2\)};
 			\vertex[below right=of e](f5){\(l_4\)};
 			\vertex[below right=of c](f6);
 			\diagram*{
 				(i1) -- [fermion] (a) -- [fermion] (f1),
 				 (a) -- [boson] (b) -- [boson, edge label'=\(V_1\)] (d),
 				 (c) -- [boson] (b) -- [boson, edge label'=\(V_2\)] (e),
 				 (f2) -- [fermion] (d),
 				 (d) -- [fermion] (f3),
 				 (f4) -- [fermion] (e),
 				 (e) -- [fermion] (f5),
 				 (i2) -- [fermion] (c) -- [fermion] (f6),
 	};
 		\end{feynman}
 	\end{tikzpicture}
	\caption*{double-resonant}
\end{subfigure}
	\caption{Categories of diagrams contributing to vector boson production processes with vector bosons $V_1$, $V_2$ decaying into leptons $l_1$, $l_2$ or $l_3$, $l_4$.}
	\label{Fig.:Diagram categories}
\end{figure}

The second difficulty arises if more complicated processes are considered involving at least two intermediate bosons. Not all diagrams participating in VB pair production processes exclusively contain final state leptonic lines connected to a single VB each, e.g.\ double-resonant diagrams for processes with two intermediate VBs as exemplified in Fig.~\ref{Fig.:Diagram categories}. For not fully-resonant diagrams, e.g.\ non- and single-resonant diagrams in the boson pair case and denoted ``non-resonant contributions'' in the following, the definition of polarisation for all intermediate VBs is unfeasible, since they cannot be interpreted as VB production times decay. 
Simply ignoring these diagrams would break electroweak gauge invariance. Thus, suitable approximations are necessary which allow to omit these diagrams while retaining gauge invariance at the same time. Two common approximations which fulfill these requirements are described below.
\begin{itemize}
	\item \textbf{Narrow-Width Approximation (NWA)} is the approximation utilised by \Sherpa and \Madgraph. It mainly replaces the denominator of the propagator by a delta function such that only on-shell contributions remain:
	\begin{equation}
		\frac{1}{q^2-m_V^2+i\Gamma_V m_V} \rightarrow \frac{\pi}{M_V \Gamma_V} \delta(q^2-M_V^2)\,.
		\label{Gl.: Narrow-Width approximation}
	\end{equation}
	\item \textbf{(Double\footnote{The pole approximation is also called double-pole approximation for the case of VB pair production processes.}-)Pole Approximation (DPA)} is used by several MC event generators such as \Phantom~\cite{Ballestrero2009}. It partially considers off-shell effects by only projecting the numerator of the propagator to on-shell momenta while leaving the denominator unchanged. This projection is not unique since sending the intermediate VBs to mass-shell requires at least the adjustment of its decay products' momenta. Further details about this approximation can be found, e.g., in Ref.~\cite{Ballestrero2017}. In the literature, the DPA is often also referred to as On-Shell projection technique (OSP).
\end{itemize}
The accuracy of both approximations is of order $\mathcal{O}(\Gamma_V/M_V)$ \cite{Ballestrero2019, Pellen2022}.
The final amplitudes result from inserting the completeness relation into the VB propagators and applying one of the approximations above, factorising them into production and decay matrix elements of external, on-shell VBs $\mathcal{M}_{\lambda_1, ..., \lambda_n}^{\mathcal{P}}$ and $\mathcal{M}^{\mathcal{D}}_{\lambda_i}$:
\begin{equation}
	\begin{aligned}
		\mathcal{M}_{\text{approx}} &\propto \sum_{\lambda_1 ... \lambda_n} \mathcal{M}_{\mu_1 ...\mu_n}^{\text{prod}} \varepsilon^{\ast \mu_1}_{\lambda_1} \cdot \cdot \cdot \varepsilon^{\ast \mu_n}_{\lambda_n} \varepsilon^{\nu_1}_{\lambda_1} \cdot \cdot \cdot \varepsilon^{\nu_n}_{\lambda_n} \mathcal{M}^{\text{decay}}_{\nu_1} \cdot \cdot \cdot \mathcal{M}^{\text{decay}}_{\nu_n} = \sum_{\lambda_1 ... \lambda_n} \mathcal{M}_{\lambda_1 ... \lambda_n}^{\mathcal{P}} \mathcal{M}^{\mathcal{D}}_{\lambda_1} \cdot \cdot \cdot \mathcal{M}^{\mathcal{D}}_{\lambda_n} \,. 
	\end{aligned}
	\label{Gl.: factorized matrix element, approximation}
\end{equation}
If both factors are gauge invariant, this also holds for the overall approximated matrix element. Since the VBs are on-shell in the polarisation dependent part of the matrix element in both approximations, the unphysical auxiliary polarisation does not contribute so that each term of the polarisation sum in Eq.~\eqref{Gl.: factorized matrix element, approximation} provides a physical polarisation contribution to the process of interest.
	
	\section{Simulation of polarised cross sections} \label{Sec.: Polarised cross sections with Sherpa}
\subsection{Basic concepts}\label{Sec.: Basic concepts Sherpa implementation}
In order to simulate polarised cross sections, amplitudes containing different VB polarisations need to be separated as detailed in the previous section. In \Sherpa, such amplitudes are available as intermediate results within the framework for simulating heavy resonances in the NWA~\cite{Hoeche2014a}. For that, the VB are produced on-shell and are subsequently decayed by an implementation of the spin-correlation algorithm introduced in Ref.~\cite{Richardson2001}. In order to at least partly recover kinematic off-shell effects of the total process, a smearing of the intermediate VB's invariant mass according to its Breit-Wigner distribution is performed after their generation, affecting only the kinematics of the final state particles. The matrix elements remain unchanged, i.e., calculated in NWA with on-shell VBs.

The main input for the spin correlation algorithm is an amplitude tensor containing all production matrix elements as a function of the VB polarisations (production tensor), $|\mathcal{M}^{\mathcal{P}}|^2_{\lambda_1\ldots\lambda_n,\lambda^\prime_1\ldots\lambda^\prime_n}$. During the simulation of the decay (cascade), a decay matrix $D_{\lambda_i\lambda^\prime_i}$ is calculated for each particle. Following from Eq.\ \eqref{Gl.: factorized matrix element, approximation}, they are defined through
\begin{equation}
	\begin{aligned}
		|\mathcal{M^{\mathcal{F}}}|_{\lambda_1\ldots\lambda_n,
		                          \lambda^\prime_1\ldots\lambda^\prime_n}^2
		&=
		\mathcal{M}^{\mathcal{P}}_{\lambda_1\ldots\lambda_n}
		\mathcal{M}^{\ast\mathcal{P}}_{\lambda^\prime_1\ldots\lambda^\prime_n}
		\mathcal{M}^{\mathcal{D}}_{\lambda_1}\mathcal{M}^{\ast\mathcal{D}}_{\lambda^\prime_1}
		\cdots
		\mathcal{M}^{\mathcal{D}}_{\lambda_n}\mathcal{M}^{\ast\mathcal{D}}_{\lambda^\prime_n}
		=
		N\cdot
		|\mathcal{M}_\text{pol}^{\mathcal{P}}|^2_{\lambda_1\ldots\lambda_n,\lambda^\prime_1\ldots\lambda^\prime_n}
		\prod_{i=1}^n D_{\lambda_i\lambda^\prime_i} \,.
	\end{aligned}
	\label{Eq.: Polarised matrix element}
\end{equation} 
The normalisation constant $N$ is the product of the normalisation constants of all decay matrices  $N=\prod_{i=1}^n n_i$ \\ with $n_i=	\mathcal{M}^{\mathcal{D}}_{\lambda_i; \rho_1 \ldots \rho_m}\mathcal{M}^{\ast\mathcal{D}}_{\lambda^\prime_i; \rho_1^\prime \ldots \rho_m^\prime} \prod_{j=1}^{m} D^j_{\rho_j, \rho^\prime_j}$ being the normalisation constant of the decay matrix of the ith~VB and $ D^j_{\rho_j, \rho\prime_j}$ denoting the decay matrices of the $m$ decay products of the respective VB. The constant $n_i$ ensures that the trace of each decay matrix $D_{\lambda_i\lambda^\prime_i}$ is one.\footnote{%
  In cases where the decay products of VB $i$ decay further, the decay
  tensor $D_{\lambda_i\lambda^\prime_i}$ is replaced (iteratively in case of longer decay cascades), up to a normalisation constant, by $D_{\lambda_i\lambda^\prime_i;\rho_1\ldots\rho_n,\rho_1^\prime\ldots\rho_m^\prime} \prod_{j=1}^m D_{\rho_j\rho_j^\prime}^j$ with $\rho_j^{(\prime)}$ being the polarisation indices of the initial VB's decay products.
}

The spin basis used for matrix element calculation in \Sherpa's built-in matrix element generator \Comix (default polarisation basis) is not the helicity basis, which is typically assumed for VB polarisation measurements, but uses constant universal reference vectors. Furthermore, other reference systems than the laboratory frame used per default may be interesting, e.g.\ to maximise the longitudinal contribution of the VBs. Hence, a transformation of polarisation definitions from one frame to another is needed.

Such a change amounts to a change of basis in the polarisation definitions, and,
hence, polarisation objects (spinors, polarisation vectors) defined in one basis can be expressed as a linear combination of polarisation objects obtained in another basis. By replacing the polarisation objects defined in the desired polarisation definition in basis $\tilde{A}$ in the corresponding matrix elements by the linear combination of the polarisation objects in basis $A$ used in the matrix element calculation, the following transformation is obtained
\begin{equation}
	\left.|\mathcal{M}|^2_{\lambda_1...\lambda_n, \lambda^\prime_1...\lambda^\prime_n}\right|_{\tilde{A}}
	=
	\sum_{\kappa_1...\kappa_n, \kappa^\prime_1...\kappa^\prime_n}
	a^{\pi_1}_{\lambda_1,\kappa_1}
	a^{\ast\pi_1}_{\lambda^\prime_1,\kappa^\prime_1}
	\cdots
	a^{\pi_n}_{\lambda_n, \kappa_n}a^{\ast\pi_n}_{\lambda^\prime_n, \kappa^\prime_n}
	\left.|\mathcal{M}|^2_{\kappa_1...\kappa_n, \kappa^\prime_1...\kappa^\prime_n}\right|_{A}\;,
	\label{Eq.: Matrixelement transformation}
\end{equation}
with $\lambda_i$ ($\kappa_i$) describing the polarisations of the $i$th particle $\pi_i$ in the desired (default) polarisation definitions.
The transformation coefficient $a^{\pi_i}_{\lambda_i, \kappa_i}$ is the linear combination coefficient for polarisation $\kappa_i$ within the polarisation object in the matrix element with polarisation $\lambda_i$. The $a^{\ast\pi_i}_{\lambda^\prime_i, \kappa^\prime_i}$ are the corresponding linear combination coefficients for the complex conjugate matrix elements. |$\mathcal{M}|^2_{\kappa_1...\kappa_n, \kappa^\prime_1...\kappa^\prime_n}$ denotes either the production tensor or a decay matrix, in the respective polarisation basis. The linear combination coefficients can be determined by solving the system of equation
\begin{equation}
	\underbrace{\begin{pmatrix}
		\tilde{\varepsilon}^1_+ & \tilde{\varepsilon}^1_- & \tilde{\varepsilon}^1_0\\
		\tilde{\varepsilon}^2_+ & \tilde{\varepsilon}^2_-& \tilde{\varepsilon}^2_0\\
		\tilde{\varepsilon}^3_+ & \tilde{\varepsilon}^3_- & \tilde{\varepsilon}^3_0\\
	\end{pmatrix}}_{\text{\parbox{3.3cm}{\centering%
		polarisation vectors in \\[-4pt] polarisation basis $\tilde{A}$}}} =
	\underbrace{\begin{pmatrix}
		\varepsilon^1_+ & \varepsilon^1_- & \varepsilon^1_0\\
		\varepsilon^2_+ & \varepsilon^2_- & \varepsilon^2_0\\
		\varepsilon^3_+ & \varepsilon^3_- & \varepsilon^3_0\\
	\end{pmatrix}}_{\text{\parbox{3.3cm}{\centering%
			polarisation vectors in \\[-4pt] polarisation basis $A$}}}
	\underbrace{\begin{pmatrix}
		a_{++} & a_{+-} & a_{+0}\\
		a_{-+} & a_{--} & a_{-0}\\
		a_{0+} & a_{0-} & a_{00}\\ 
	\end{pmatrix}}_{\text{\parbox{3.3cm}{\centering%
		linear combination \\[-4pt] coefficients}}},
	\label{Eq.:system of equation linear combination coefficients} 
\end{equation} by inversion. For the inverse of a (3x3) matrix an analytical formula exists, so no numerical determination is necessary. Note that the zeroth component of the polarisation vector is omitted in Eq.~\eqref{Eq.:system of equation linear combination coefficients} since not all components of a polarisation vector are independent according to Eq.~\eqref{Eq.: Polarisation vector properties}.

\subsection{Structure of the new polarisation framework}
Following the ideas and concepts introduced in the previous section, the event generation for processes with heavy resonances in NWA remains unchanged if polarised cross sections should be simulated. Only copies of the production tensor and the decay matrices are made before they are contracted with each other to result in the unpolarised cross section in NWA. In practical terms, the polarised cross sections are provided simultaneously via additional event weights in each event. Those are calculated directly from the matrix elements by
\begin{enumerate}
\item \textbf{Transformation} of the polarised production tensor and decay matrices to get matrix elements defined in the polarisation basis of interest according to Eq.~\eqref{Eq.: Matrixelement transformation}
\item \textbf{Multiplying} production tensor and decay matrices according to Eq.~\eqref{Eq.: Polarised matrix element}
\item \textbf{Labelling/Identification} of tensor entries according to the polarisation combinations of interest
\item \textbf{Normalisation} to the tensor sum to obtain polarisation fractions
\end{enumerate}

Those fractions are than multiplied with the event cross section and stored as event weights. Details about provided weights and their naming scheme can be found in Appendix~\ref{Appendix: Provided polarisation weights}. \\
This approach enables the simulation of polarised cross sections of all possible polarisation combinations in all polarisation definitions of interest in a single simulation run. Furthermore, the interference between different polarisations can be simulated directly by summing over all off-diagonal entries of the amplitude tensor:
\begin{equation}
	\underbrace{|\mathcal{M}|^2\vphantom{\sum_{\lambda_1 \neq \lambda_1'... \lambda_n \neq \lambda_n'}}}_{\text{coherent sum}}
	=
	\underbrace{\sum_{\lambda_1...\lambda_n} |\mathcal{M}_{\lambda_1...\lambda_n}^\mathcal{F}|^2\vphantom{\sum_{\lambda_1 \neq \lambda_1'... \lambda_n \neq \lambda_n'}}}_{\text{polarised contributions}, \, \text{incoherent sum}}
	+
	\underbrace{\sum_{\lambda_1 \neq \lambda_1'... \lambda_n \neq \lambda_n'} \mathcal{M}_{\lambda_1...\lambda_n}^\mathcal{F}  \mathcal{M}_{\lambda_1^\prime...\lambda_n^\prime}^{\ast \mathcal{F}}}_{\text{interference}}\,,
	\label{Gl.: Interference term general}
\end{equation}
Thus, an interference template can be directly provided for polarisation analyses. It can be included as an additional background in polarisation measurements without relying on histogram subtraction methods that can lead to large statistical uncertainties.

All common reference systems for polarisation definitions are supported, including the laboratory, the parton-parton and the center-of-mass frame of all intermediate particles. Due to the a posteriori approach of adjusting the polarisation definition in the matrix elements an extension to other reference systems (and spin bases) is straightforward. The framework can be applied for an arbitrary number of intermediate VBs.

Practical details, the user input syntax to simulate polarised cross sections with \Sherpa and a complete list of currently implemented polarisation definitions can be found in a dedicated section of the \Sherpa user manual~\cite{SherpaMasterManual}. A short depiction of the input syntax is also given in Appendix~\ref{Appendix: Input structure}.

	\subsection{Calculation of polarised cross sections at nLO QCD with \Sherpa} \label{Sec.: NLO pol Sherpa}
As described in Sec.~\ref{Sec.: Basic concepts Sherpa implementation}, the new polarisation features are based on \Sherpa's framework for computing factorised matrix elements of the production and decay of unstable, intermediate particles. This is not limited to LO: The production matrix elements can also include NLO contributions which can in turn be used for the polarisation calculation enabling the simulation of higher-order polarisation effects. We limit ourselves to NLO QCD corrections.

The simulation of NLO polarisation effects currently relies on some approximations depending on the event type. As a consequence, while unpolarised NLO+PS matched calculations with \Sherpa's \SMCatNLO method~\cite{Hoeche2011} retain their complete NLO accuracy, the polarisation fractions computed with the following construction can only approximate it. We construct the amplitude tensor $|\mathcal{M}|^2_{\kappa_1...\kappa_n, \kappa^\prime_1...\kappa^\prime_n}$ depending on the event type.
\begin{description}
  \item[\quad$\boldsymbol{\mathbb{H}}$ events.]
    This category comprises both hard well-separated emissions beyond the parton shower starting scale, and process-specific corrections to the universal soft-collinear emission pattern below it. The corresponding amplitude tensor is constructed from the real emission amplitude itself and thus contains all necessary information in the hard-emission regime.
  \item[\quad$\boldsymbol{\mathbb{S}}$ events, resolved emission.]
    This category comprises the universal soft-collinear radiation pattern the parton shower approximation produces above its infrared cut-off. The corresponding amplitude tensor is again constructed using the complete real emission amplitude. Hence, in combination with the treatment for $\mathbb{H}$-events in this regime the correct polarisation fractions, up to NLO, are used for both soft and hard emissions.
  \item[\quad$\boldsymbol{\mathbb{S}}$ events, unresolved emission.]
    Finally, this category contains, to NLO accuracy, all unresolved emissions, i.e.\ emissions below the parton shower infrared cutoff and all virtual corrections. The amplitude tensor is solely constructed using the Born expression and all virtual and ultra-soft and/or -collinear emission corrections thereupon are neglected. As the number of events in this category is generally small, and this construction is used to determine the polarisation fractions in the otherwise fully NLO-accurate unpolarised sample, the error introduced in this way is expected to be small.
\end{description}
This approximation generally yields satisfactory results and will be denoted as nLO in the following. We investigate it in more detail in Sec.\ \ref{sec:pheno}.

A similar phenomenon holds for merged setups of tree-level multi-leg matrix elements. In the matrix element jet production region the full real-emission kinematics is taken into account also for the polarisation calculations and thus the bulk of higher-order QCD corrections for them is included.
The same methods can not be reliably applied for NLO EW corrections, or the EW${}_\text{virt}$ approximation \cite{Kallweit2015} for that matter, because there the virtual effects in the production and decay will have a significant impact on VB polarisation fractions.

	\section{Validation of the implementation at fixed leading order} \label{Sec.: Validation}

\begin{table}[t!]
	\renewcommand{\arraystretch}{1.3}
	\centering
	\begin{small}
		\begin{tabular}{llll}
			\toprule
			Process $\mathcal{O}(\alpha^6)$ & Reference & Approximation & Reference system\\
			\midrule
			$\mathrm{W}^+$$\mathrm{W}^+$jj: pp $\rightarrow$ $\mathrm{e^+}$ $\nu_\text{e}$ $\mu^+\nu_\mu$jj & \cite{Ballestrero2020} & OSP & Lab, COM \\
			\hline
			$\mathrm{W}^+$$\mathrm{W}^-$jj: pp $\rightarrow$ $\mathrm{e^+}$ $\nu_\text{e}$ $\mu^-$ $\bar{\nu}_\mu$jj & \cite{Ballestrero2020} & OSP & Lab, COM \\
			$\mathrm{W}^+$$\mathrm{W}^-$jj: pp $\rightarrow$ $\mathrm{\mu^+}$ $\nu_\mu$ $\text{e}^-$ $\bar{\nu}_\text{e}$jj & \cite{Ballestrero2017} & OSP & Lab\\
			\hline
			$\mathrm{W}^{+}$Zjj: pp $\rightarrow$ $\mathrm{e^+}$ $\nu_\text{e}$ $\mu^+\mu^-$jj & \cite{Ballestrero2020} & OSP & Lab, COM \\
			$\mathrm{W}^{+}$Zjj: pp $\rightarrow$ $\mu^+$ $\nu_\mu$ $\text{e}^+\text{e}^-$jj & \cite{Ballestrero2019}  & OSP & Lab \\
			\hline
			ZZjj: pp $\rightarrow$ $\mathrm{e^+}$ $\mathrm{e}^-$ $\mu^+\mu^-$jj & \cite{Ballestrero2020} & OSP &  Lab, COM \\
			& \cite{Ballestrero2019} & RES NO OSP & Lab \\
			\bottomrule
		\end{tabular}
	\end{small}
	\caption[Details about the literature studies used to validate the new \Sherpa implementation for simulating polarised cross sections for vector boson production]{Details about the literature studies used to validate the new \Sherpa implementation for simulating polarised cross sections for vector boson production processes. Here, OSP denotes the On-Shell Projection technique, RES NO OSP the use of double-resonant diagrams in the full off-shell phase space, while Lab signals the use of the laboratory frame, and COM the center-of-mass frame of the VB pair, to define the polarisation states.}
	\label{Tab.:Literature}
\end{table}

The \Sherpa implementation presented in the previous section is validated by comparing polarised integrated cross sections and differential distributions obtained with the new polarisation framework with literature data for several pure electroweak VB-pair-production processes in association with two jets at fixed leading order ($\mathcal{O}(\alpha^6)$)\footnote{For simplicity, those processes are abbreviated as VBS processes in this work, even though not only VBS diagrams contribute to them.}. Tab.~\ref{Tab.:Literature} summarises all literature data which is included in this validation study. All considered literature data were obtained with \Phantom \cite{Ballestrero2009} for an LHC beam setup with 13~TeV proton-proton center-of-mass energy assuming SM dynamics, and rely on different approximations to omit single- and non-resonant diagrams than \Sherpa. Hence, literature and \Sherpa predictions may differ on the order of $\mathcal{O}(\Gamma_{\text{V}}/m_{\text{V}})$. The VB polarisation is defined in the helicity basis. Simulation parameters and phase space definitions are chosen identical to the literature, for details see also Appendix~\ref{Appendix:Simulation setup validation}. Typically, two different phase space definitions are distinguished in the literature based on whether lepton acceptance criteria are applied (fiducial setup) or not (inclusive setup).

For each investigated process, two simulation runs are done:
\begin{itemize}
	\item \textbf{Full calculation (denoted ``full''):} No approximation is applied so that all off-shell effects of the intermediate VBs as well as all effects from single- and non-resonant contributions are retained.
	\item \textbf{Polarised calculation:} \Sherpa's spin-correlated narrow-width approximation as introduced in Sec.~\ref{Sec.: Polarised cross sections with Sherpa} is used to compute polarised contributions, distinguishing longitudinal (\textbf{``L''}), transverse (\textbf{``T''}, summing left- and right-handed polarised contributions as well as left-right interference terms) and their \emph{incoherent} sum (\textbf{``polsum''}). Furthermore, the unpolarised cross section in the NWA (\textbf{``unpol''}) is calculated, which additionally also contains the interferences, i.e.\ the \emph{coherent} sum of all polarised matrix elements. The unpol differs from the full result by the missing non-resonant contributions and not-completely covered off-shell effects.
\end{itemize}
Uncertainties reported within this work only take the limited MC statistics into account. There are also systematic uncertainties due to missing higher-order contributions, which could be estimated by scale and PDF variations.

The data generated with \Sherpa is analysed via the analysis framework \Rivet~\cite{Bierlich2020}, details are given in Appendix~\ref{Appendix:Simulation setup validation}.

\begin{table}[t!]
	\renewcommand{\arraystretch}{1.3}
	\centering
	\small
	\begin{tabular}{lllll}
		\toprule
		Process & \multicolumn{2}{c}{ $\sigma_{\text{Phantom}}$ [fb]} & \multicolumn{2}{c}{ $\sigma_{\text{Sherpa}}$ [fb]} \\
		\hline
		& full & unpol & full & unpol\\
		\midrule
		$\text{W}^+\text{W}^+$jj & 3.185(3) & 3.167(2) & 3.1814(13) & 3.1839(10)\\
		$\text{W}^+\text{W}^-$jj &  4.651(2) &  4.641(2) & 4.631(22) & 4.6707(22)\\
		$\text{W}^+$Zjj & 0.5253(3) & 0.5210(3) & 0.5258(11) & 0.52471(27) \\
		ZZjj &  0.1270(1) & 0.1264(1) & 0.12715(25) &  0.12801(10) \\
		\bottomrule
	\end{tabular}
	\caption{Integrated full and unpolarised cross sections for several pure electroweak vector boson pair production processes in association with two jets at fixed LO obtained with \Phantom from Ref.~\cite{Ballestrero2020} and \Sherpa in the inclusive phase space defined in Sec.~\ref{Appendix:Simulation setup validation}}
	\label{Tab.: Full and unpolarised cross sections from Phantom and Sherpa}
\end{table}

A first comparison of integrated cross sections is summarised in Tab.~\ref{Tab.: Full and unpolarised cross sections from Phantom and Sherpa}, which compares the unpolarised predictions obtained with \Phantom in Ref.~\cite{Ballestrero2020} with those resulting from the new polarisation framework in \Sherpa in the inclusive setups. As is evident, \Sherpa's NWA leads to a very good approximation of the full result for all investigated processes and phase space regions and observed deviations are $\ll$1\%. The difference between the unpolarised results from the literature and \Sherpa amounts to 1.3\% or less which is fully within the expected accuracy of $\mathcal{O}(\Gamma/M)$~($\sim 2.5\%$).

\begin{figure}[t]
 	\includegraphics[width=\textwidth]{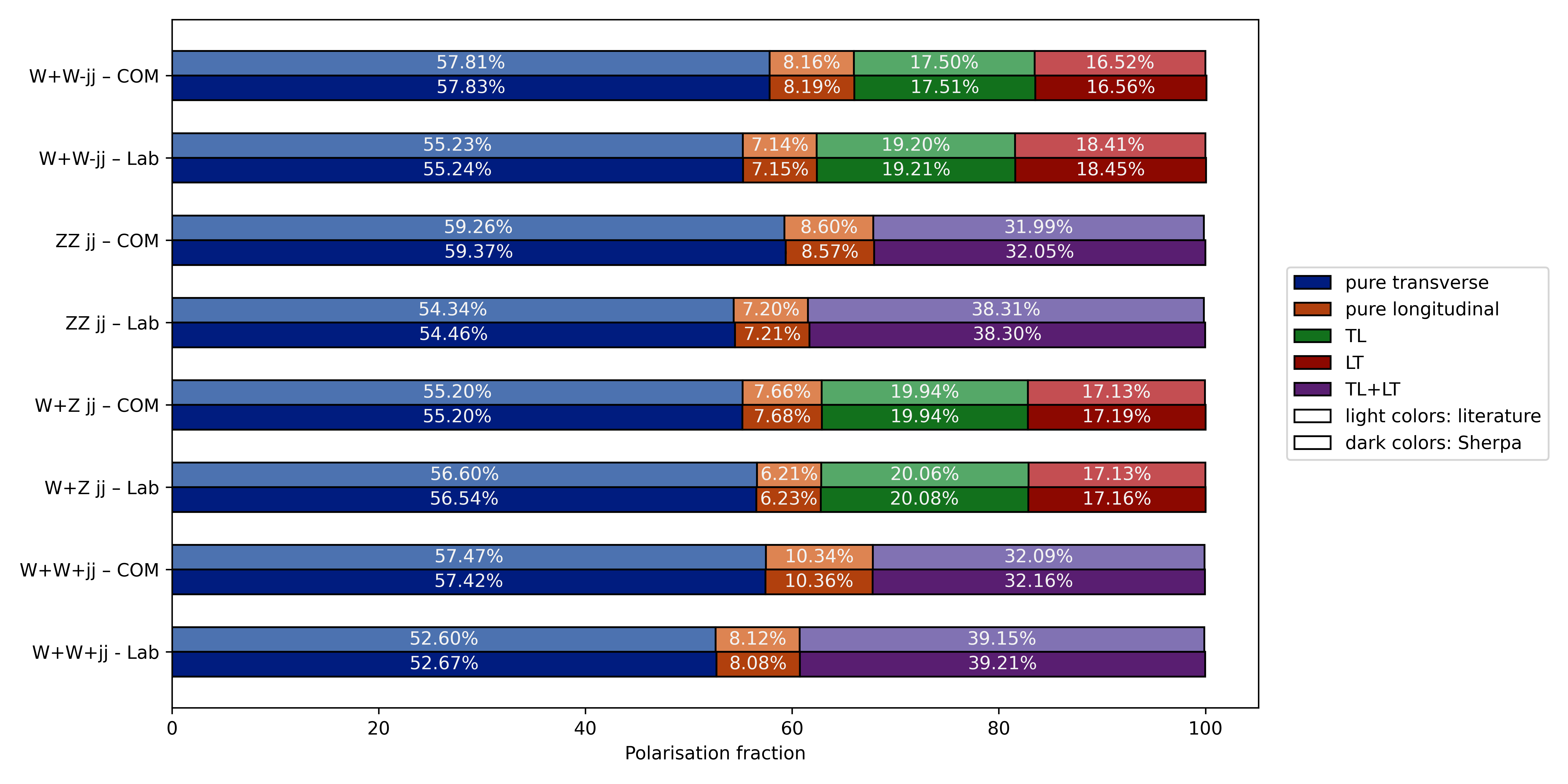}
	\caption{Comparison of polarisation fractions obtained with \Phantom in Ref.~\cite{Ballestrero2020} and \Sherpa for several pure electroweak VB pair production processes in association with two jets at fixed leading order in the inclusive phase space defined in Sec.~\ref{Appendix:Simulation setup validation}, vector boson polarisation is defined in the laboratory (Lab) or the vector boson pair center-of-mass frame (COM). Polarisation fractions are computed relative to the unpolarised results indicating vanishing interference between different polarisations as expected in the absence of lepton selection requirements. Statistical uncertainties of the polarisation fractions are of $\mathcal{O}(10^{-2})$\%.}
	\label{Fig.: Polarisation fractions VBS inclusive phase space}
\end{figure}

In Fig.~\ref{Fig.: Polarisation fractions VBS inclusive phase space}, the polarisation fractions for the investigated processes computed with \Sherpa in the inclusive setups with polarisation defined in the laboratory (Lab) and the VB pair center-of-mass (COM) frames, respectively, are displayed in comparison with the \Phantom results indicating an excellent agreement with deviations of less than 0.6\%. The simulated interference contribution is compatible with zero, as expected when no lepton acceptance criteria are applied.

\begin{figure}[t!]
	\centering
	\begin{subfigure}{0.49\textwidth}
		\begin{subfigure}{\textwidth}
			\includegraphics[width=\textwidth]{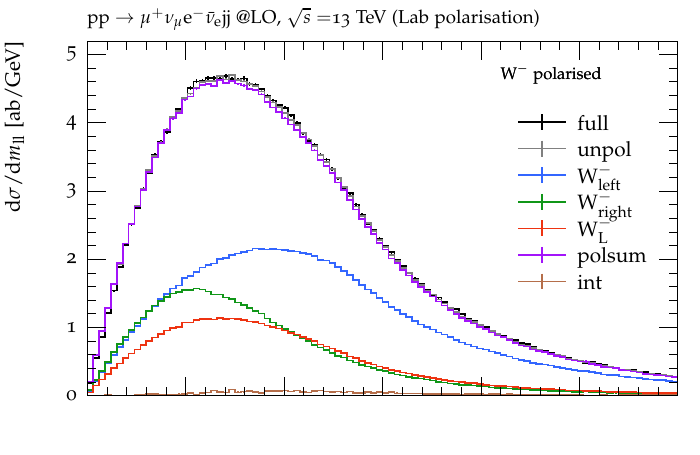}
			\vskip -36pt
			\includegraphics[width=\textwidth]{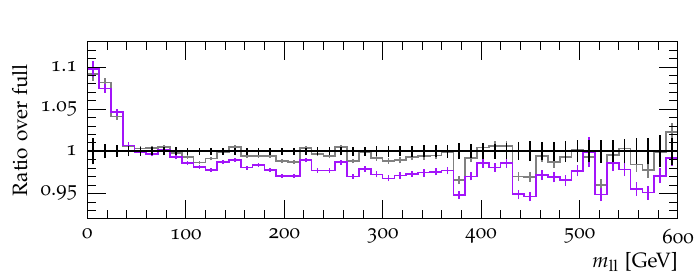}
		\end{subfigure}
	\end{subfigure}
		\begin{subfigure}{0.49\textwidth}
		\begin{subfigure}{\textwidth}
			\includegraphics[width=\textwidth]{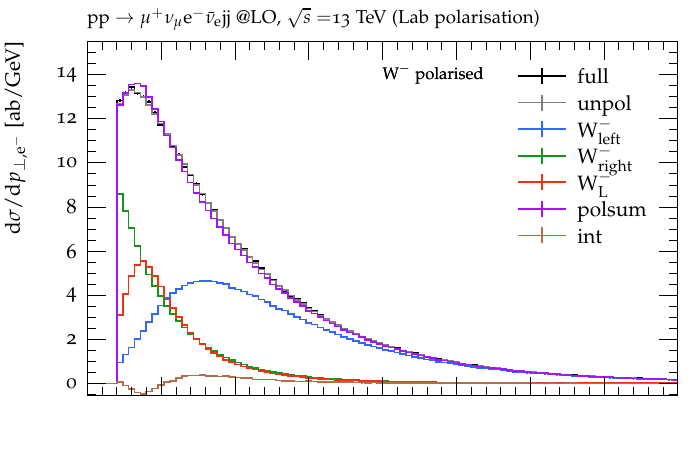}
			\vskip -36pt
			\includegraphics[width=\textwidth]{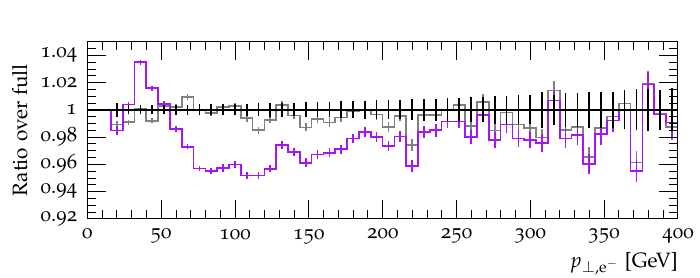}
		\end{subfigure}
	\end{subfigure}
	\caption{
	  Single-polarised distributions of the charged leptons invariant mass $m_{\text{ll}}$ (left) and the transverse momentum of the electron $p_{\perp, \text{e}^-}$ for the $\text{W}^+\text{W}^-$jj process in the fiducial phase space (details in main text) obtained with \Sherpa. The $\text{W}^+$ boson is considered unpolarised, the polarisation of the $\text{W}^-$ boson is defined in the laboratory frame (lab). The \Sherpa distributions are in very good agreement with those computed with \Phantom in Ref.~\cite{Ballestrero2017}, Fig.~6.
	}
	\label{Fig.: Validation W+W-}
\end{figure}

\begin{figure}[t!]
	\begin{subfigure}{0.49\textwidth}
		\begin{subfigure}{\textwidth}
			\includegraphics[width=\textwidth]{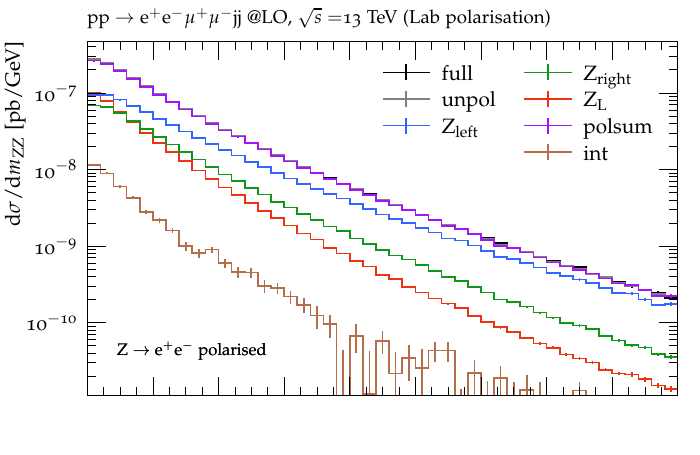}
		\end{subfigure}
		\vskip -36pt
		\begin{subfigure}{\textwidth}
			\includegraphics[width=\textwidth]{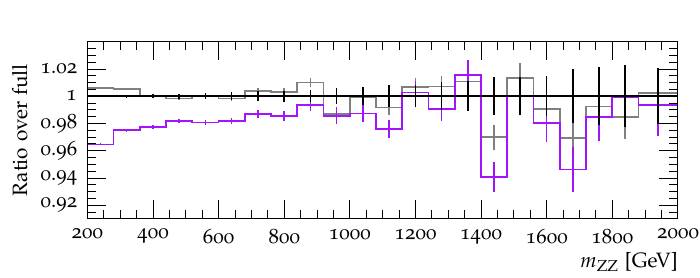}
		\end{subfigure}
	\end{subfigure}
	\begin{subfigure}{0.49\textwidth}
		\includegraphics[width=\textwidth]{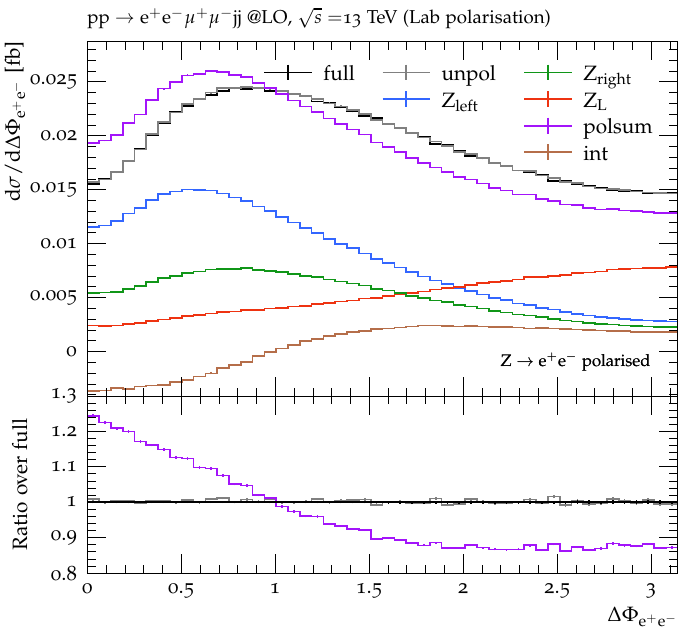}
	\end{subfigure}
	\caption{
	  Single-polarised distributions of the ZZ-invariant mass $m_{\text{ZZ}}$ (left) and the the azimuthal separation $\Delta\Phi_{\text{e}^+\text{e}^-}$ (right) for the ZZjj process in the fiducial phase space (details in main text) obtained with \Sherpa. The Z boson decaying into $\mu^+\mu^-$ is considered as unpolarised, the polarisation of the Z boson decaying into $\text{e}^+\text{e}^-$ is defined in the laboratory frame (lab). The ZZ-invariant mass distributions are also computed with \Phantom in Ref.~\cite{Ballestrero2019}, Fig.~4, the \Sherpa distributions shown here are in very good agreement with those. 
	}
	\label{Fig.: Validation ZZ}
\end{figure}

\begin{figure}[t!]
	\centering
	\begin{subfigure}{0.49\textwidth}
		\includegraphics[width=\textwidth]{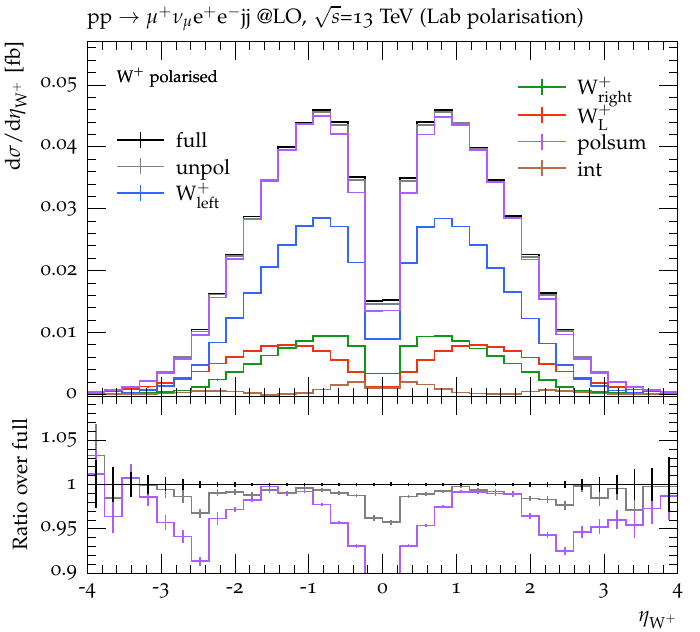}
	\end{subfigure}
	\begin{subfigure}{0.49\textwidth}
		\includegraphics[width=\textwidth]{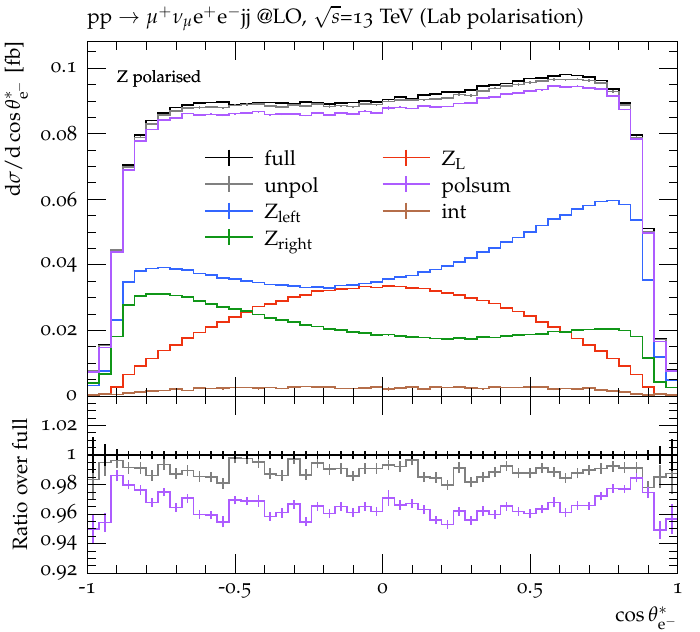}
	\end{subfigure}
	\caption{
	  Single-polarised distributions of the $\text{W}^+$ boson pseudorapidity $\eta_{\text{W}^+}$ (left) and the electron decay angle $\cos{\theta_{\text{e}^-}}$ (right) for the $\text{W}^+\text{Z}$jj process in the fiducial phase space (details in main text) obtained with \Sherpa. The $\text{Z}$ boson is considered as unpolarised in the left figure, the $\text{W}^+$ boson in the right figure, the polarisation of the respective polarised boson is defined in the laboratory frame (lab). The \Sherpa distributions are in very good agreement with those computed with \Phantom in Ref.~\cite{Ballestrero2019}, Fig.~7.
	}
	\label{Fig.: Validation W+Z}
\end{figure}

\begin{figure}[t!]
	\centering
	\begin{subfigure}{0.49\textwidth}
		\includegraphics[width=0.94\textwidth]{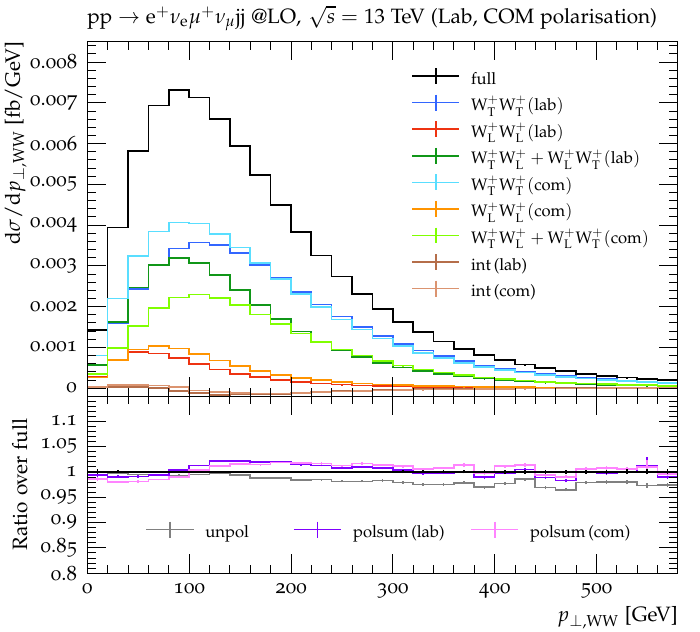}
	\end{subfigure}
	\begin{subfigure}{0.49\textwidth}
		\includegraphics[width=0.94\textwidth]{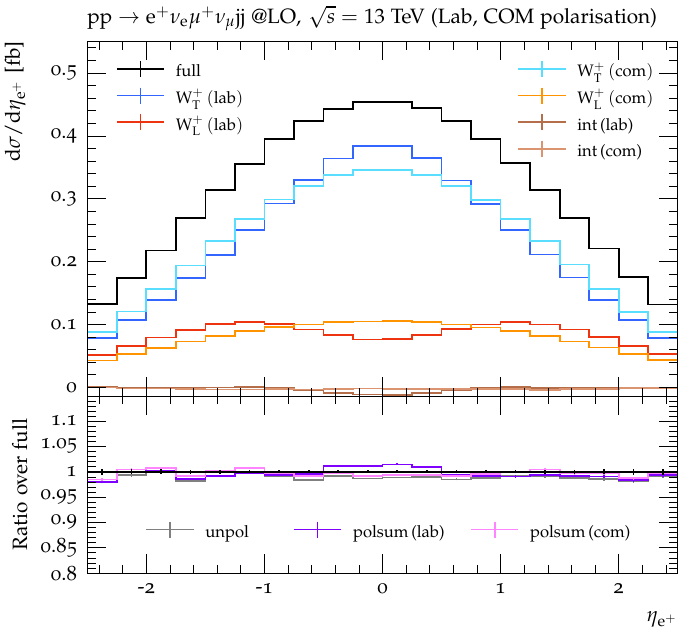}
	\end{subfigure}
	\caption{
	  Double-polarised distributions of the $\text{W}^+$$\text{W}^+$ transverse momentum $p_{\perp, \text{WW}}$ (left) and single-polarised distributions of the positron pseudorapidity $\eta_{\text{e}^+}$ (right) for the $\text{W}^+ \text{W}^+$jj process in the fiducial phase space (details in main text) obtained with \Sherpa. Polarised differential distributions and the ratios over the full off-shell result are shown in the top respectively bottom graph of each subfigure. For the single-polarised distributions, the $\text{W}^+$ boson decaying into an anti-muon-neutrino pair is considered as unpolarised. Polarisations of both $\text{W}^+$ bosons are defined in the $\text{W}^+$$\text{W}^+$-center-of-mass frame (com) and the laboratory frame (lab). The \Sherpa distributions are in very good agreement with those computed with \Phantom in Ref.~\cite{Ballestrero2020}, Fig.~2 and 3. 
	}
	\label{Fig.: Validation W+W+}
\end{figure}

For a more differential validation a selection of the observables investigated in Refs.\ \cite{Ballestrero2017, Ballestrero2019, Ballestrero2020} is reproduced with our new implementation in \Sherpa in Figs.~\ref{Fig.: Validation W+W-}-\ref{Fig.: Validation W+W+}. The (partially) polarised differential cross sections are computed in the fiducial phase spaces defined in Appendix~\ref{Appendix:Simulation setup validation}. Resulting integrated cross sections are displayed in Tabs.~\ref{Tab.:ZZjj-validation-total cross section fiducial phase space}-\ref{Tab.: ssWW validation fiducial phase space} in the Appendix~\ref{Appendix.: Additional validation results} for completeness. They agree within the expected accuracy (deviations <1.5\%). Except the $\text{W}^+\text{W}^+$jj process, where also results for the polarisation defined in the COM and double-polarised differential cross sections are discussed in the literature, all distributions are computed in the Lab and are single-polarised, i.e.\ only the polarisations of VBs decaying in the electron channel are considered.

Generally, excellent agreement between \Phantom and \Sherpa calculations is found. Observed deviations are small, far below the expected accuracy due to the different approximations used. Furthermore, it is confirmed that the unpolarised predictions obtained by applying \Sherpa's spin-correlated NWA reproduce the full result at a comparable level as the DPA for the investigated processes and phase spaces.

The interference contribution, which, at variance with the literature, our implementation allows to compute directly, has also been added to the presented figures. For most observables and phase space regions studied here the interference is small in comparison to other contributions, amounting to a few percent at most. However, there are also observables where the interference contribution can reach a similar size than the longitudinal contribution in some phase space regions (e.g. around $\eta_{\text{W}^+}=0$ in the $\text{W}^+$Zjj process in Fig.~\ref{Fig.: Validation W+Z}) or can even be very large over large phase space regions (Fig.~\ref{Fig.: Validation ZZ}). It is here, where the direct access to this component facilitates precise measurements of polarised cross sections at any collider experiment.

	\section{Polarised cross sections beyond leading order}
\label{sec:pheno}

The validation study presented in the previous section reveals a very good agreement between the new \Sherpa polarisation framework and the literature for several investigated VB production processes at fixed LO, leading to the conclusion that the new implementation is functional.
The inclusion of higher perturbative orders in the hard matrix elements and multijet-merged calculations provide two ways to obtain higher-order corrections to simulated cross sections. As outlined in Sec.~\ref{Sec.: NLO pol Sherpa}, it is possible to apply the new polarisation framework in \Sherpa's NLO calculations. The current implementation, however, leads to polarised predictions at nLO accuracy only, since the calculation of polarisation fractions does not take loop corrections into account. Therefore, the first part of this section focuses on testing, whether the approximation introduced in Sec.~\ref{Sec.: NLO pol Sherpa} can nonetheless deliver reliable predictions of the main complete NLO polarisation behaviour. This is done by comparing \Sherpa results with full fixed-order NLO calculations from the literature, Ref.~\cite{Denner2021}, where predictions with \MoCaNLO{}+\Recola{}+\Collier~\cite{Actis2016,Denner2016} in the DPA are presented.
This will be followed by a discussion of the polarised cross sections obtained by combining \Sherpa's merging method with the new polarisation framework. In particular, the influence of different merging scales is investigated.
We choose inclusive production of a $\text{W}^+$Z~boson pair with fully leptonic decays as our testbed.

\subsection{Simulation setup}
\label{Sec.: Simulation setup WZ nLO}

In order to study higher-order predictions with \Sherpa in inclusive $\text{W}^+$Z production, simulation settings and the phase space are chosen according to Ref.~\cite{Denner2021}. This allows the comparison with full NLO fixed-order results published there. More specifically, the {pp $\rightarrow$ $\text{e}^+\, \nu_\text{e} \, \mu^+ \, \mu^-+X$} process is studied for an LHC setup with 13~TeV proton-proton center-of-mass energy assuming SM dynamics and using the simulation settings as well as phase space definitions given in Tab.~\ref{Tab.: simulation settings, PS for NLO WZ, Wj}. The polarisation is defined in the helicity basis. All leptons and quarks except the top quark are considered to be massless.

\textbf{\Sherpa simulations at nLO+PS accuracy} use Born-level matrix elements calculated with \Amegic \cite{Krauss2002}, except for the amplitudes employed for the polarisation fraction calculations, the VB decays and calculations at LO. Those amplitudes as well as real corrections are computed with \Comix \cite{Gleisberg2008}. Virtual corrections are provided by OpenLoops \cite{Buccioni2019} using the \Sherpa interface to this program.
These simulations showcase a simulation of events including parton shower effects, which allows to interface hadronization models and thus provide a fully-realistic simulation. Since these effects are not taken into account in the reference predictions, one can expect differences in the comparison in particular in regions which become sensitive to resummation. The matching of the fixed-order NLO calculation to the resummation of the parton shower is done by \Sherpa's internal implementation of the \MCatNLO method \cite{Hoeche2011}. The core-scale of the processes is set by \Sherpa's METS-scale setter \cite{Hoeche2009}.

\textbf{Merged calculations with \Sherpa} in this section investigate the {$\text{pp}\rightarrow$ $\text{e}^+\, \nu_\text{e} \, \mu^+ \, \mu^-$} process merged with {$\text{pp}\rightarrow$ $\text{e}^+\, \nu_\text{e} \, \mu^+ \, \mu^-$j} at LO. All matrix elements are provided by \Comix. The merging algorithm implemented in \Sherpa is an extension of the CKKW method \cite{Catani2001} as detailed in Ref.~\cite{Hoeche2009} (called MEPS@LO). The merging scale is varied in an extreme range between 20 and 1000 GeV for instructive purposes.

All \Sherpa simulations performed in this section apply \Sherpa's default shower simulation~\cite{Schumann2007}, QED radiation~\cite{Schoenherr2008}, hadronisation~\cite{Chahal2022}, hadron decays and multiple interactions \cite{Bothmann2019,SherpaMasterManual}. QCD jets are identified by the anti-$k_t$ algorithm~\cite{Cacciari2008} with a jet resolution parameter of $R=0.4$.

The simulation data is analysed with a \Rivet analysis based on the ATLAS analysis of Ref.\ \cite{ATLASCollaboration2019}. 
Selection criteria applied during the event generation are set more inclusive than in Tab.~\ref{Tab.: simulation settings, PS for NLO WZ, Wj}. The more stringent selection criteria detailed in Tab.~\ref{Tab.: simulation settings, PS for NLO WZ, Wj} are then applied on the reconstructed final state jets and dressed final state leptons at the analysis stage.
We then focus on double-polarised cross sections.

\begin{table}[t!]
	\renewcommand{\arraystretch}{1.3}
	\centering
	\small
	\begin{tabular}{ll}
		\toprule
		PDF-Set from LHAPDF6 \cite{Buckley2015}  & NNPDF31\_nlo\_as\_0118 \cite{NNPDFCollaboration2017} \\
		Electroweak scheme $G_\mu$ &  $G_\mu=1.16638\cdot 10^{-5}\, \text{GeV}^{-2}$ and complex mass scheme (full), \\
		& real EW parameters (polarisation) \\
		Strong coupling $\alpha_S(M_\text{Z})$ & 0.118\\
		Core-Scale & $\mu$=$\frac{1}{2}(M_{\text{Z}}+M_{\text{W}})$\\
		VB pol masses &$M_\text{W}=80.352 \text{ GeV}$, $M_\text{Z}=91.153 \text{ GeV}$\\
		VB pol widths & zero for all VBs in polarisation calculations \cite{Ballestrero2017}, \\
		&$\Gamma_\text{W}=2.084 \text{ GeV}$,  $\Gamma_\text{Z}=2.4943 \text{ GeV}$ otherwise\\
		\hline
		Phase space &  $p_{\perp, \text{e}^+} >$ 20 GeV, $p_{\perp, \mu^\pm} >$ 15 GeV, $|y_\text{l}|<2.5$ \\
		& $\Delta R_{\mu^+ \mu^-} > 0.2$, $\Delta R_{\mu^\pm \text{e}^+}>0.3$ \\ 
		& 81 GeV < $M_{\mu^+ \mu^-}$ < 101 GeV, $M_{T, \text{W}}$  >30 GeV\\
		\bottomrule
	\end{tabular}	
	\caption{
	  Settings and phase space definition used for the inclusive $\text{W}^+$Z production at nLO+PS and LO+1j merged calculations.
	  The transverse mass is defined as $M_{T, \text{W}}=\sqrt{2p_{\perp, \text{e}^+}p_{\perp,\nu_{\text{e}}}(1-\cos \Delta \phi_{\text{e}^+\nu_{\text{e}}})}$.
	}
	\label{Tab.: simulation settings, PS for NLO WZ, Wj}

\end{table}

\subsection{Polarised cross sections at nLO+PS}
\label{Sec. nLOPS}

\begin{small}
	\begin{table}[t!]
		\renewcommand{\arraystretch}{1.3}
		\centering
		\begin{tabular}{lllllll}
			\toprule
			$\text{W}^+\text{Z}$   & $\sigma^{\text{NLO}}$ [fb]   & Fraction [\%] & K-factor  & $\sigma^{\text{nLO+PS}}_{\text{\Sherpa}}$ [fb]   & Fraction [\%]  & K-factor \\
			\midrule
			full                   & 35.27(1)                       & &1.81            & 33.80(4)                                &    &         \\
			unpol                  & 34.63(1)                       & 100    &  1.81     & 33.457(26)                      & 100        & 1.79\\
			\hline
			\multicolumn{7}{c}{Laboratory frame}\\
			\hline
			L-U & 8.160(2) & 23.563(9) & 1.93 & 7.962(5) & 23.796(25)& 1.91\\ 
			T-U &26.394(9)& 76.217(34) & 1.78 & 25.432(21) & 76.01(9)& 1.75\\
			int & 0.066(10) (diff) & 0.191(29) & 2.00 & 0.064(7)   & 0.191(22)& 2.40(40)\\
			\hline
			U-L & 9.550(4) & 27.577(14) & 1.73 & 9.275(16) & 27.72(5) & 1.72 \\
			U-T & 25.052(8) & 72.342(31) & 1.83 & 24.156(18) & 72.20(8) &1.81 \\ 
			int &0.028(10) (diff) & 0.081(29) & -0.49 & 0.026(7) & 0.079(22) & -0.471(34) \\
			\hline
			L-L                    & 2.063(1)                       & 5.9573(33)  & 1.91 & 2.0128(18)                      & 6.016(7)    & 1.90 \\
			L-T                    & 6.108(2)                       &  17.638(8) & 1.93  & 5.958(5)                        & 17.807(20) & 1.91\\
			T-L                    & 7.409(4)                       & 21.395(13)  & 1.69 &  7.185(12)                       & 21.47(4)   & 1.68\\
			T-T                    & 18.964(7)                      & 54.762(26)  & 1.80 &18.215(16)                      & 54.44(6)    &1.77 \\
			int                    & 0.086(13) (diff)               & 0.248(35)   & -2.97  & 0.087(7)                       & 0.259(20) &   -2.7(4) \\ 
			\hline
			\multicolumn{7}{c}{$\text{W}^+$Z-center-of-mass-frame}\\ 
			\hline
			L-U & 7.308(2) & 21.103(8) & 2.09 & 7.132(5) & 21.316(22) & 2.08\\
			T-U & 27.14(1) & 78.371(37) & 1.75 & 26.153(17) & 78.17(8) & 1.73\\
			int & 0.182(10) & 0.526(29) & 1.28 &0.173(10) &0.516(30)& 1.30(4)\\
			\hline
			U-L & 7.137(2) & 20.609(8) & 2.07 & 6.976(5) & 20.850(22) & 2.07\\
			U-T & 27.449(9) & 79.264(35) & 1.75 & 26.441(26) & 79.03(10) & 1.72\\
			int & 0.044(10) (diff) & 0.127(29) & $\infty$ & 0.041(4) & 0.122(13) & -15(8)\\
			\hline
			L-L                    & 1.968(1)                       & 5.6829(33) & 1.31 & 1.9018(19)                      & 5.684(7)  & 1.28\\
			L-T                    & 5.354(1)                       & 15.461(5)   &2.65 & 5.241(4)                        & 15.665(17)   & 2.65\\
			T-L                    & 5.097(2)                       & 14.718(7)  & 2.68 & 5.002(4)                       &  14.951(16)  & 2.69\\
			T-T                    & 21.992(9)                      & 63.506(32)  &1.62 & 21.098(16)                        & 63.06(7)  & 1.59\\
			int                    & 0.219(13) (diff)                               &  0.632(38)           & 1.54 & 0.215(9)  &  0.641(26)  &  1.65(5)\\
			\bottomrule
			\end{tabular}
		\caption{
		  Integrated single- and double-polarised cross sections for the inclusive $\text{W}^+$Z production including higher order QCD corrections in the phase space defined in Tab.~\ref{Tab.: simulation settings, PS for NLO WZ, Wj} obtained with \MoCaNLO{}+\Recola{}+\Collier~\cite{Actis2016,Denner2016} in Ref.~\cite{Denner2021} (full fixed-order NLO correction) and \Sherpa (nLO+PS) for polarisations defined in the laboratory and $\text{W}^+$Z-center-of-mass frame; polarisation fractions are calculated relative to the unpolarised result, K-factors are obtained by dividing the NLO(nLO+PS) cross sections by the LO (LO+PS) ones which lead to statistical errors of $\mathcal{O}(10^{-3})$, except for the interference ($\mathcal{O}(10^{-1})$).
		}
		\label{Tab.:WZ inclusive-validation-total cross section nlo}
	\end{table}
\end{small}

Tab.~\ref{Tab.:WZ inclusive-validation-total cross section nlo} compares the integrated polarised cross sections obtained with \Sherpa's nLO+PS polarisation setup with the literature results of Ref.~\cite{Denner2021} containing all NLO corrections. For completeness, also single-polarised cross sections are displayed. Both the laboratory (Lab) as well as the $\text{W}^+$Z-center-of-mass reference frame (COM) are investigated. NLO corrections in $\text{W}^+$Z production can be very large with K-factors around two. This is due to both new flavour channels opening at NLO and the approximate radiation amplitude zero, present at Born level, being filled by NLO real-emission corrections  \cite{Denner2021}.

Comparing the unpolarised \Sherpa result with the literature results, while full agreement was observed at LO, the unpolarised \Sherpa cross section is about 3.4\% smaller than the literature value. This discrepancy is caused by parton shower effects which are accounted for in the \Sherpa calculation but not in the fixed-order reference result. The \Sherpa unpolarised result approximates the full cross section better than 1.5\%, implying that \Sherpa's NWA provides a good approximation to the full result also at NLO.

More interestingly, the polarisation fractions show an agreement of better than 1.5\%. This is smaller than the expected accuracy due to the different on-shell approximations used in the two calculations, and the inherent approximation of the nLO calculation used in \Sherpa. This leads to the conclusion that virtual corrections do not seem to have a significant influence on the polarisation fractions and \Sherpa's nLO setup captures the relevant part of the NLO corrections. It is particularly promising that the differing behaviour in both reference frames is reproduced accurately.

\begin{figure}[t] 
	\centering
	\begin{subfigure}{0.4\textwidth}
		\centering
		\includegraphics[width=0.9\textwidth]{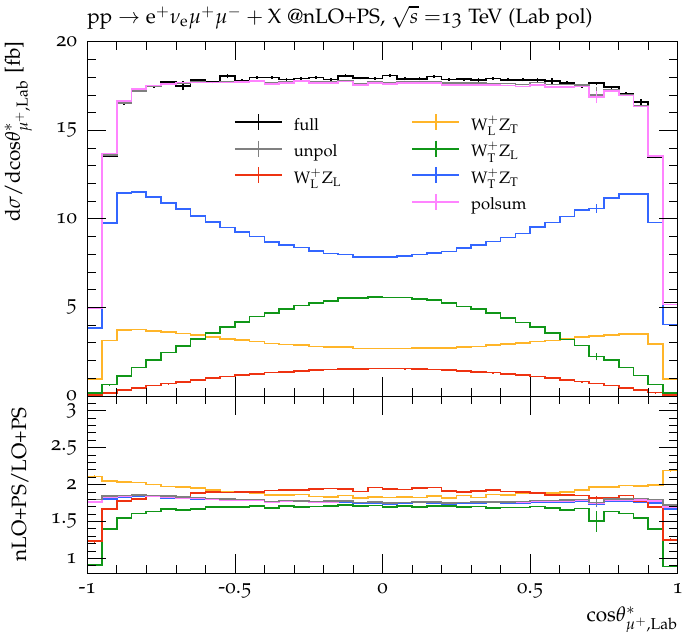}
	\end{subfigure}
	\hspace*{0.1\textwidth}
	\begin{subfigure}{0.4\textwidth}
		\centering
		\includegraphics[width=0.9\textwidth]{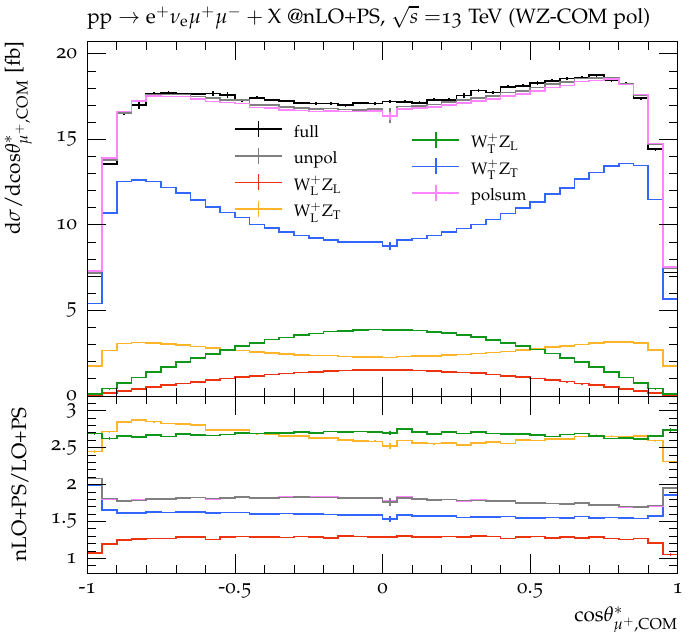}
	\end{subfigure}\\[2mm]
	\begin{subfigure}{0.4\textwidth}
		\centering
		\includegraphics[width=0.9\textwidth]{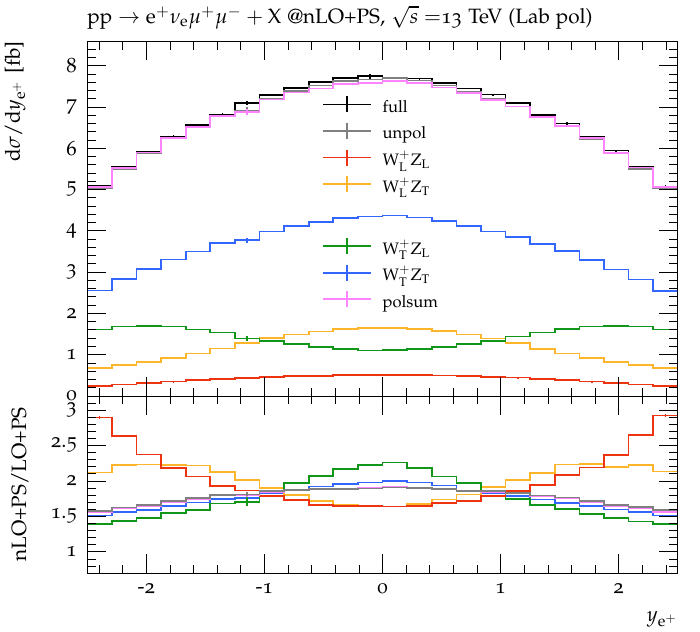}
	\end{subfigure}
	\hspace*{0.1\textwidth}
	\begin{subfigure}{0.4\textwidth}
		\centering
		\includegraphics[width=0.9\textwidth]{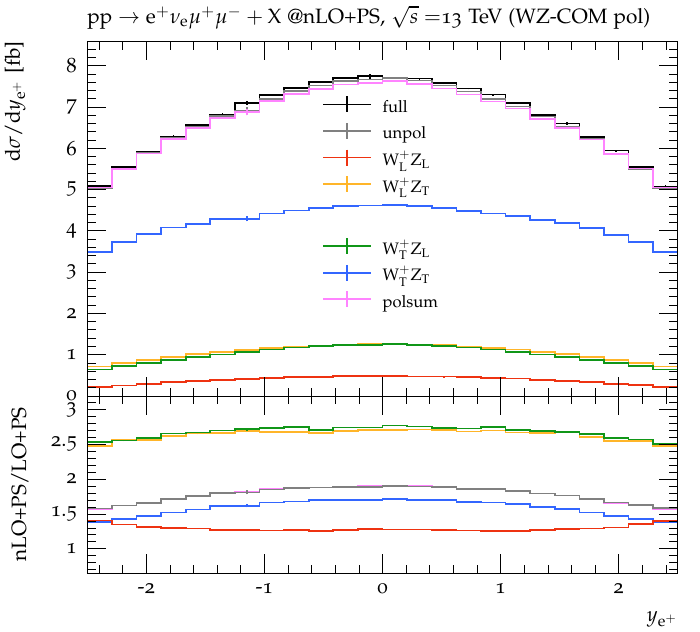}
	\end{subfigure}
	\caption{
	  Double-polarised distributions of the anti-muon decay angle $\cos \theta^\ast_{\mu^+}$ (top) and the positron rapidity $y_{\textbf{e}^+}$ (bottom) in inclusive $\text{W}^+$Z production obtained with \Sherpa at nLO+PS (polarised distribution) / NLO+PS (unpol, full) accuracy; polarisation states are defined in the laboratory (Lab, left) and the $\text{W}^+$Z center-of-mass frame (WZ-COM, right), K-factors are the ratio of n(N)LO+PS over LO+PS cross sections. All \Sherpa distributions agree well with MOCANLO+COLLIER+RECOLA~\cite{Actis2016,Denner2016} full NLO fixed-order predictions in Ref.~\cite{Denner2021}, Fig.~1 and 3.
	}
	\label{Fig.: WZ nLO+PS validation}
\end{figure}

\begin{figure}[t!]
	\centering
	\begin{subfigure}{0.4\textwidth}
		\centering
		\includegraphics[width=0.9\textwidth]{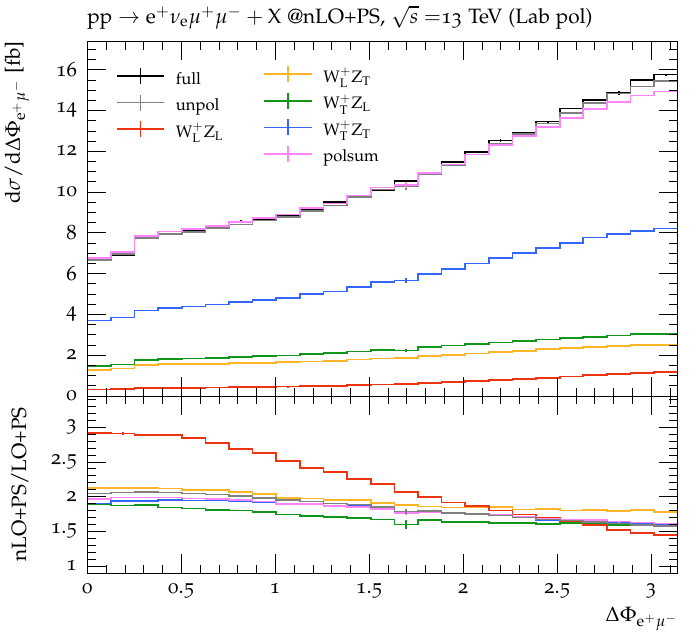}
	\end{subfigure}
	\hspace*{0.1\textwidth}
	\begin{subfigure}{0.4\textwidth}
		\centering
		\includegraphics[width=0.9\textwidth]{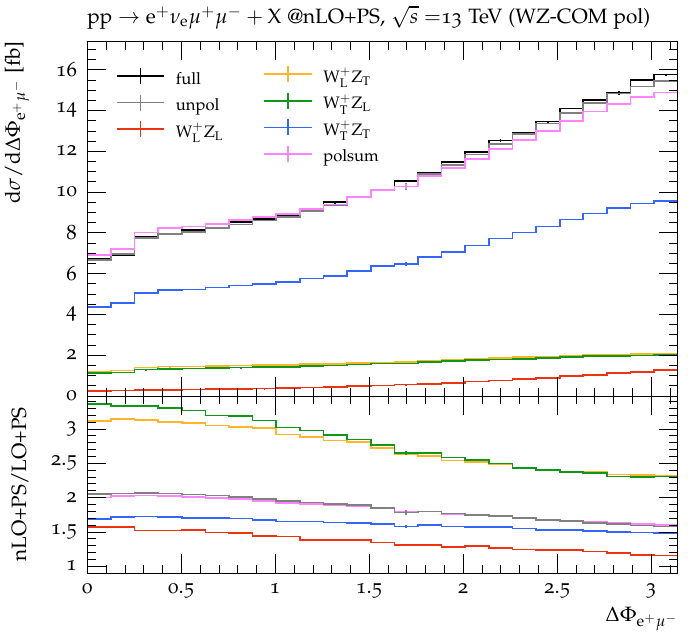}
	\end{subfigure}\\[2mm]
	\begin{subfigure}{0.4\textwidth}
		\centering
		\includegraphics[width=0.9\textwidth]{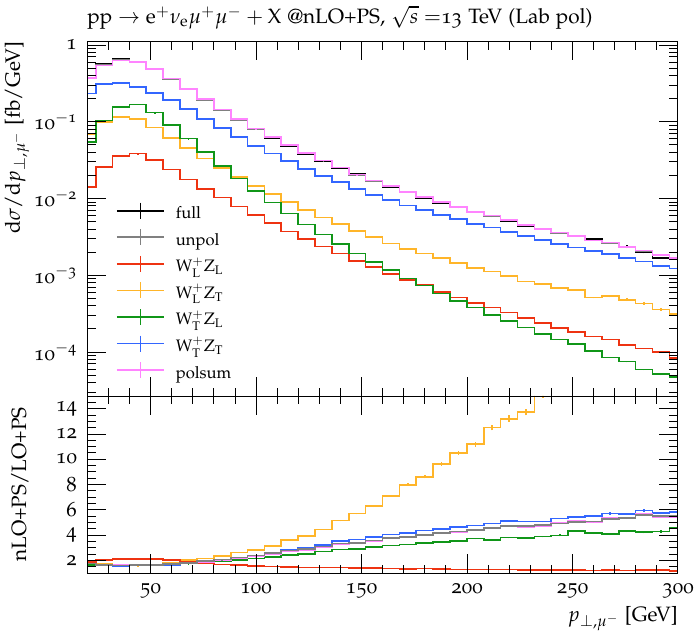}
	\end{subfigure}
	\hspace*{0.1\textwidth}
	\begin{subfigure}{0.4\textwidth}
		\centering
		\includegraphics[width=0.9\textwidth]{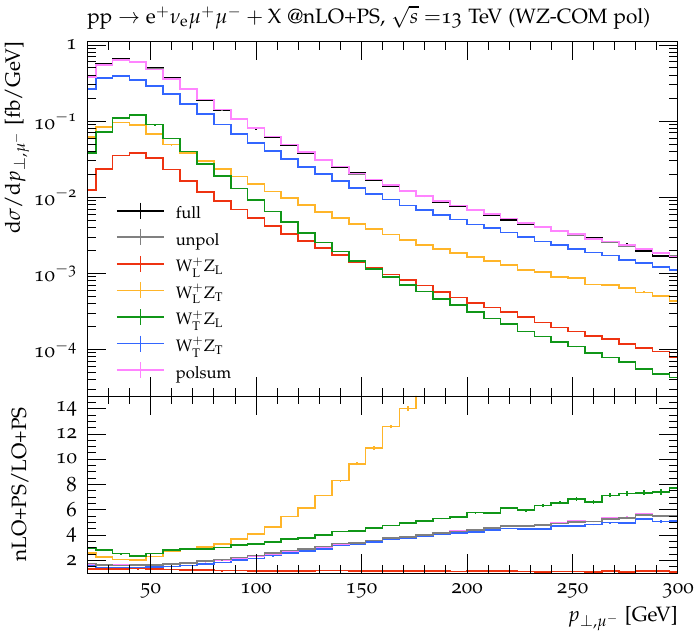}
	\end{subfigure}
	\caption{
	  Double-polarised distributions of the azimuthal angle between the positron and the muon $\Delta \Phi_{e^+\mu^-}$ (top) and the muon transverse momentum $p_{\perp,\mu^-}$ (bottom) in inclusive $\text{W}^+$Z production. Details as in Fig.\ \ref{Fig.: WZ nLO+PS validation}. The $\Delta \Phi_{e^+\mu^-}$ distributions are also calculated with MOCANLO+COLLIER+RECOLA~\cite{Actis2016,Denner2016} at full NLO fixed-order in Ref.~\cite{Denner2021}, Fig.~5. The \Sherpa results shown here agree well with these literature results. 
	}
	\label{Fig.: WZ nLO+PS validation continued}
\end{figure}

Figs.~\ref{Fig.: WZ nLO+PS validation} and \ref{Fig.: WZ nLO+PS validation continued} show the distributions for the decay angle of the anti-muon~$\cos \theta^\ast_{\mu^+}$, the positron rapidity~$y_{\text{e}^+}$, the azimuthal angle between the positron and the muon~$\Delta \Phi_{e^+\mu^-}$ and the muon transverse momentum~$p_{\perp,\mu^-}$ obtained with \Sherpa for the Lab and COM polarisation definition.
The unpolarised (unpol) distributions approximate the full predictions (full) well, mostly showing deviations of about 2-3\% or less. Hence, off-shell and
non-factorisable effects are small.

The NLO corrections show a strong sensitivity to the polarisation states under consideration, and can be very non-uniform across a given observable, e.g.\  $y_{\text{e}^+}$ with polarisation states defined in the Lab frame. Additionally, shape and size of the NLO corrections for a certain polarisation state can strongly depend on the frame in which the respective polarisation states are defined. Here, the distributions of $\cos\theta_{\mu^+}^*$  or, again, $y_{\text{e}^+}$ are poignant examples.

The shapes of the K-factors, however, exhibit some deviations from the results of Ref.\ \cite{Denner2021} for the polarisation definition in the laboratory frame. For $|y_{\text{e}^+}|>2$, the differential K-factor for the LL contribution is increased by up to 7\% compared to the literature.
A similar deviation can be seen in the differential K-factors for $\cos\theta_{\mu^+}^*$. For $|\cos \theta^\ast_{\mu^+}|>0.85$ the K-factor of the LL (TL) contributions decreases from $K\approx 1.8\;(1.6)$ to values around 1 in the first and last bin, an effect not seen in the literature. A comparison of results at fixed LO with LO+PS predictions reveals that effects are induced by higher-order corrections effected by the parton shower. It raises the cross section for the LL and TL distributions in the phase space regions near $\theta^*$ of 0 and $\pi$, respectively. The shower in the LO+PS simulation thus already covers parts of the NLO corrections in those phase space regions. Hence,  the K-factor, and thus the remaining NLO correction, is decreasing as a consequence. It is interesting to note that a similar effect can be seen for the longitudinal polarisations of the W boson in $\cos\theta_{e^+}^*$.
The $\cos\theta_{\mu^+}^*$ distributions for polarisations defined in the COM, however, do not show such a large growth in LO+PS distributions compared to the fixed LO results. This is consistent with the observation that deviations in the K-factors only occur for the Lab decay angles.

All in all, the \Sherpa nLO+PS calculation, despite its reduced formal accuracy, can reproduce the shapes and K-factors of Ref.\ \cite{Denner2021} for all investigated observables excellently, with the added benefit of being matched to parton shower evolution and fully differential event simulation.

\FloatBarrier
\subsection{Polarised cross sections in multijet-merged calculations}
\label{Sec.: MEPS}

\begin{table}[t!]
	\renewcommand{\arraystretch}{1.3}
	\centering
	\small
	\begin{tabular}{lllllllll}
	\toprule
	$\text{W}^+\text{Z}$ & $\sigma^{\text{nLO+PS}}_{\text{\Sherpa}}$ [fb] & K & $\sigma^{Q_c=20\text{GeV}}_{\text{\Sherpa}}$ [fb]   & K& $\sigma^{Q_c=40\text{GeV}}_{\text{\Sherpa}}$ [fb]    & K&  $\sigma^{Q_c=80\text{GeV}}_{\text{\Sherpa}}$ [fb]  & K    \\
	\midrule
	full                   &    33.95(4)                            &             & 29.633(28)  && 29.311(20) &&&\\
	unpol               & 33.439(24) & 1.78  & 29.357(10)                     & 1.56      & 29.124(10)                      &  1.55  & 27.194(9) & 1.45  \\
	\hline
	int                    & 0.235(10) & 1.72 & 0.178(5)                       & 1.31   & 0.166(4)                        & 1.21 & 0.117(4) & 0.86   \\
	L-L                    &1.891(7) & 1.28 & 1.6656(15)                     & 1.12   & 1.6686(14)                      & 1.13 & 1.6095(14) & 1.09 \\
	L-T                    & 5.231(7) & 2.61 &  4.6227(29)                       & 2.31  & 4.4568(27)                        & 2.22 & 3.9203(23)   &1.96\\
	T-L                    & 5.007(6) & 2.66 &  4.3994(25)                       & 2.34  & 4.2086(23)                        & 2.24 & 3.6560(21) & 1.94\\
	T-T                    & 21.074(17) & 1.58 & 18.491(8)                     & 1.39    & 18.625(7)                      & 1.40   & 17.890(7) & 1.35 \\
	\hline
	\vspace{-3mm}
	&&&&&&&&\\
	\vspace{0.6mm}
	& $\sigma^{Q_c=200\text{GeV}}_{\text{\Sherpa}}$ [fb]   & K &
	$\sigma^{Q_c=500\text{GeV}}_{\text{\Sherpa}}$ [fb]  & K & $\sigma^{Q_c=1000\text{GeV}}_{\text{\Sherpa}}$ [fb] & K & $\sigma^{\text{LO+PS}}_{\text{\Sherpa}}$ [fb] &   \\
	\hline
	unpol           & 22.221(13)& 1.18  &  19.324(11) & 1.03 & 18.870(11) & 1.00 & 18.803(10) & \\
	\hline
	int                & 0.115(6) & 0.84 & 0.121(5) & 0.89 & 0.136(5) & 1.00 & 0.137(4) &  \\
	L-L                    & 1.5052(22)& 1.02 & 1.4871(21) & 1.00 & 1.4857(21) & 1.00 & 1.4823(18) &  \\
	L-T                    & 2.7834(30)& 1.39 & 2.1168(23)& 1.06 & 2.0122(21) &1.00 & 2.0041(18) &\\
	T-L                     & 2.5805(27)& 1.37 & 1.9824(21)& 1.05 & 1.8926(20) & 1.01 & 1.8821(17) &\\
	T-T                   & 15.236(10)& 1.15 & 13.616(9)& 1.02 & 13.343(9) & 1.00 & 13.298(8) &\\
	\bottomrule
\end{tabular}
	\caption{
	  Integrated double-polarised cross sections for the simulation of inclusive $\text{W}^+$Z production with up to one jet merged to parton shower at LO. Polarised cross sections and K-factors (K) for different merging scales $Q_c$ are shown in comparison with nLO+PS and LO+PS results. Polarisation states are defined in the $\text{W}^+$Z-center-of-mass frame (COM), K-factors are obtained from dividing the nLO+PS and LO+1j cross sections by the LO+PS cross section, statistical errors on the K-factors are of the order of $\mathcal{O}(10^{-3})$, except for the interference ($\mathcal{O}(10^{-1})$).
	}
	\label{Tab.: Results from merged setup WZ production COM}
\end{table}

\begin{figure}[t!]
 	\includegraphics[width=\textwidth]{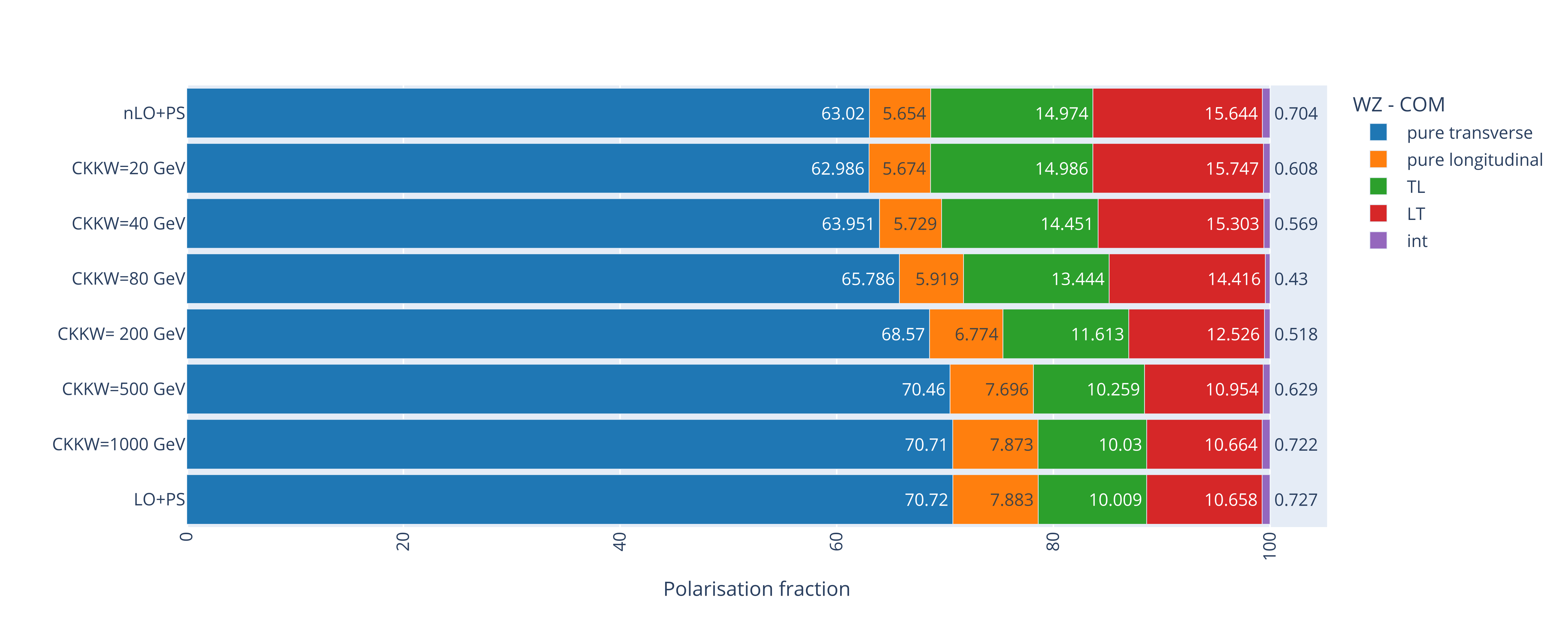}
	\caption{
	  $\text{W}^+$Z center-of-mass frame (COM) polarisation fractions for inclusive $\text{W}^+$Z production at nLO+PS, LO+PS and LO+1j merged with PS with different CKKW-merging scales $Q_c$ between 20 and 1000~GeV. Polarisation fractions are calculated relative to the unpolarised result from data given in Tab.~\ref{Tab.: Results from merged setup WZ production COM}.
	}
	\label{Fig.: Polarisation fractions inclusice WZ merged simulations COM}
\end{figure}

In this section we study the combination of multijet-merged calculations with the new polarisation framework in \Sherpa, again using the example of inclusive $\text{W}^+\text{Z}$ production. We investigate whether the dependency of the polarised cross sections on the merging scale matches the expectations, i.e.
\begin{description}
	\item[Small merging scales.] The resolved (according to the merging scale) one-jet emission corrections to the parton shower are applied over large parts of the phase space, similar to in an NLO-matched calculation. Hence, in cases where the exact virtual correction in an NLO-matched calculation has negligible impact on polarisation correlations, i.e.\ where our above nLO-matched approximation is valid, we expect to recover very similar polarisation fractions in this case.
	\item[Large merging scales.] The resolved (according to the merging scale) one-jet emission corrections to the parton shower are applied over only a small fraction of phase space. We thus expect to recover polarisation fractions more and more similar to those at LO.
\end{description}
Tab.~\ref{Tab.: Results from merged setup WZ production COM} and Fig.~\ref{Fig.: Polarisation fractions inclusice WZ merged simulations COM} summarise the integrated cross sections and polarisation fractions obtained from these merged calculations with different merging scales applied and polarisation defined in the $\text{W}^+$Z-COM frame. Results for LO+PS and nLO+PS simulations are given for comparison.
Comparing the off-shell (full) and unpolarised on-shell (unpol) calculations in the nLO+PS and the multijet-merged setup with the smallest merging scale investigated, a deviation of 15\% for the total cross section is mainly caused by the missing virtual corrections in the merged results. All polarised cross sections and associated K factors are reduced accordingly. Conversely, the polarisation fractions when calculated using a reasonably low merging scale of $Q_c=20\,\text{GeV}$ agree remarkably well with the nLO+PS result at the sub-percent level or better. Hence, the missing finite real-emission corrections to the parton shower for emissions with scales below 20~GeV have no relevant effect on the polarisation fractions, in line with the soft-collinear factorisation of the amplitudes.
With increasing merging scale, a continuous change of cross sections, K-factors and polarisation fractions towards the LO results can be observed, reaching it
for $Q_c\gtrsim 500\,\text{GeV}$.
It is interesting to note that not all polarisation fractions are equally affected by the absence (or presence) of hard-emission corrections in the different multijet-merged calculations. While, e.g., the fraction of the cross section carried by the interference of different polarisation states is reduced by 7\% when raising $Q_c$ from 20 to 40\,GeV, the fraction of the cross section wherein both VBs are polarised longitudinally remains approximately constant.
In summary, the use of merging scales $Q_c>40\,\text{GeV}$ is counter-indicated if real-emission effects are considered to be important to accurately describe the polarisation fractions in a given sample. We will thus restrict our discussion in the remainder of this section to reasonable merging scales $Q_c\le 40\,\text{GeV}$.

\begin{figure}[t!]
	\centering
	\begin{subfigure}{0.49\textwidth}
		\includegraphics[width=\textwidth]{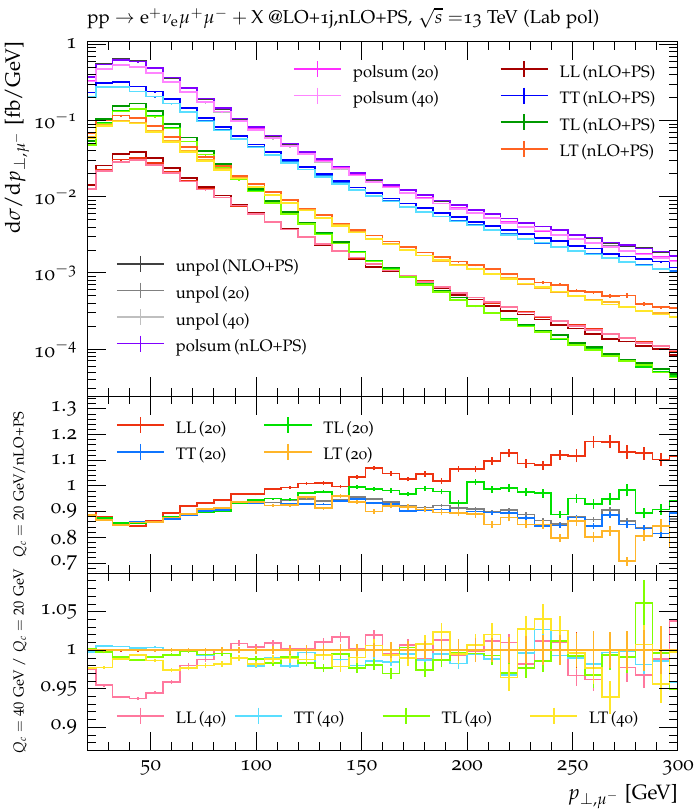}
		\centering
	\end{subfigure}
	\begin{subfigure}{0.49\textwidth}
 	    \includegraphics[width=\textwidth]{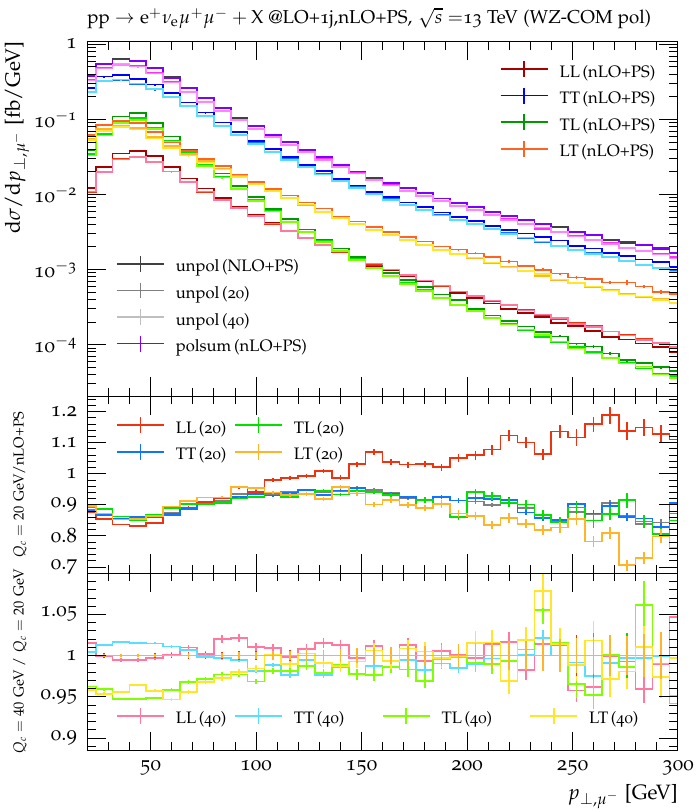}
    	\centering
	\end{subfigure}\\
	\begin{subfigure}{0.49\textwidth}
		\includegraphics[width=\textwidth]{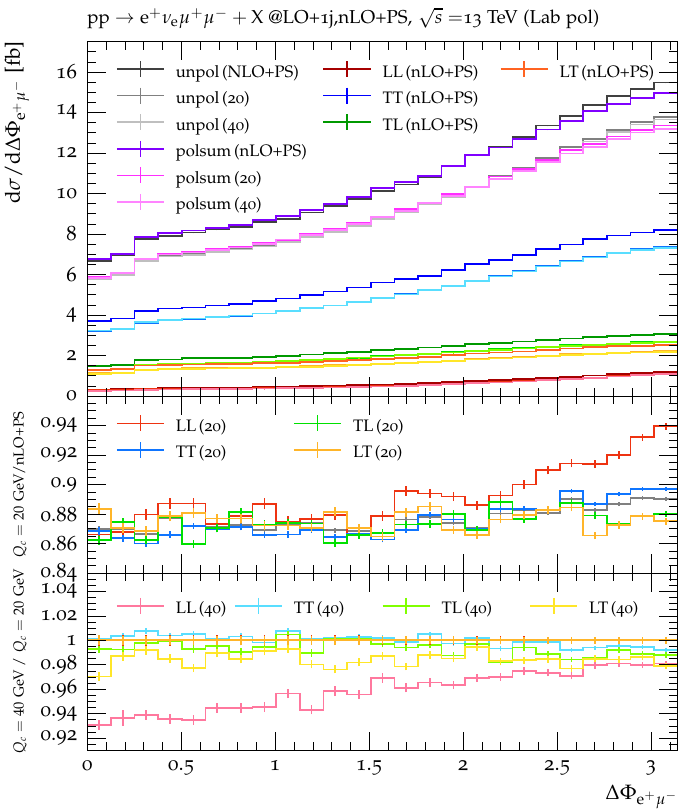}
		\centering
	\end{subfigure}
	\begin{subfigure}{0.49\textwidth}
		\includegraphics[width=\textwidth]{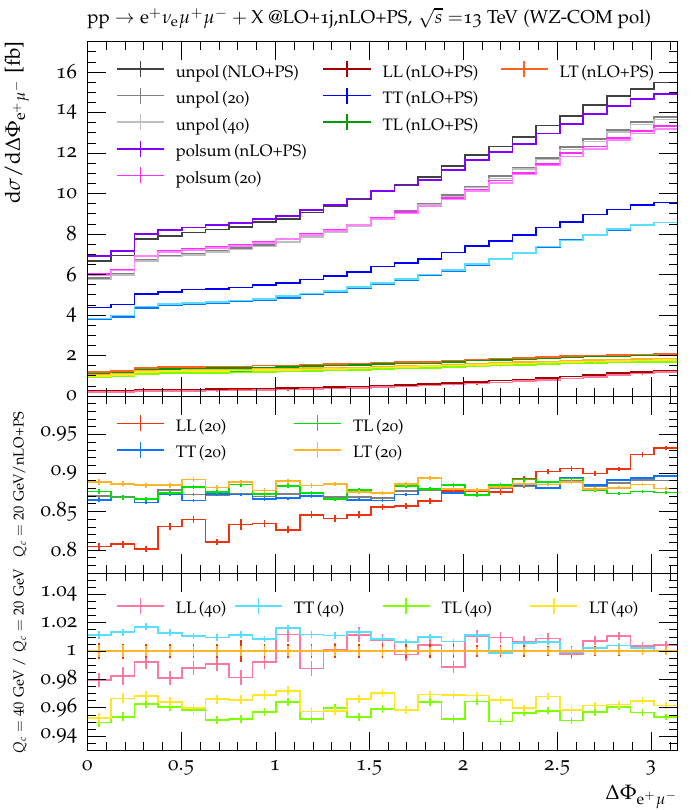}
		\centering
	\end{subfigure}
	\caption{
	  Double-polarised distributions of the muon transverse momentum $p_{\perp, \mu^-}$ (top) and the azimuthal separation of the positron and the muon $\Delta \Phi_{\text{e}^+\mu^-}$ (bottom) for the inclusive $\text{W}^+$Z-production process. \Sherpa results obtained with LO+1j merged calculations with CKKW-merging scales of 20 and 40 GeV are shown in comparison with results at nLO+PS accuracy. Polarisation states are defined in the laboratory frame (left) and the $\text{W}^+$Z-center-of-mass frame (right).
	}
	\label{Fig.: WZ MEPS pT dphi}
\end{figure}

We now turn to examine differential distributions.
Fig.~\ref{Fig.: WZ MEPS pT dphi} displays the polarised distributions in the muon transverse momentum, $p_{\perp, \mu^-}$, and the azimuthal separation between positron and muon, $\Delta \Phi_{\text{e}^+ \mu^-}$, obtained at nLO+PS and from merged calculations with reasonably small merging scales (20 and 40\,GeV) to illustrate the influence of merged calculations on distributions w.r.t.\ the nLO+PS result (upper ratio plot) and the effect of a merging scale variation (lower ratio plot). The ratios of the polarised contributions are calculated with respect to their counterparts in the nLO+PS/$Q_c=20$~GeV simulations.

The missing virtual and exact real-emission corrections for very soft-collinear emission at scales below 20\,GeV reduce all merged distributions by roughly 15\% w.r.t.\ the nLO+PS results\footnote{We remind the reader that our nLO+PS results are in fact NLO accurate for unpolarised observables.}, as already observed for the integrated results. Their influence is, however, not constant over the entire investigated phase space. This can be seen in particular for the muon transverse momentum. Similarly, the impact of the missing virtual and very soft-collinear real-emission corrections also depends on the polarisation state under consideration and the frame they are defined in. In particular, the LL component reveals a markedly different behaviour compared to all other polarisation combinations. Not only does this polarisation combination exhibit
a smaller impact of these missing contributions w.r.t. the nLO+PS result at
large $p_{\perp, \mu^-}$, $\Delta \Phi_{\text{e}^+ \mu^-}$, the lepton azimuthal separation $\Delta \Phi_{\text{e}^+ \mu^-}$ at small angles also shows a frame dependent effect of these contributions.
In either case, though, all mismodeling effects w.r.t\ the nLO+PS result are contained within $\approx 5\%$ of the flat K-factor hypothesis for the dominant polarisation combination and only reach a deviation of up to 10\% (20\%) for $\Delta \Phi_{\text{e}^+ \mu^-}$ ($p_{\perp, \mu^-}$) for the strongly suppressed LL contribution.

As a consequence of the above observations, the merging scale variation (lower ratio plot in Fig.~\ref{Fig.: WZ MEPS pT dphi}), varying the amount of exact real-emission corrections effected onto the parton shower, also affects different polarisation states and definitions differently. For the polarisation states defined in the laboratory frame, the missing real-emission corrections from the one-jet matrix element between 20 and 40~GeV only have a significant influence on the LL component, again, for $\Delta \Phi_{\text{e}^+ \mu^-}$<~2.5 and $p_{\perp, \mu^-}$<~80~GeV, while, conversely, for the COM definition they only have a noticeable impact on mixed polarised cross sections (LT and TL) in the whole $\Delta \Phi_{\text{e}^+ \mu^-}$ phase space and for $p_{\perp, \mu^-}$<~100~GeV.
For these polarisation combinations an accurate description of emissions at moderate scales, $20-40\,\text{GeV}$, is vital. The TT polarisation combinations in both the Lab and COM frame definitions, however, remain nearly unaffected by the merging scale variation with deviations of 2\% or less.

Connecting these merging scale dependences with the K-factor dependence of the
different polarisation states we find some correlation between the two. In phase space regions where polarisation states are affected by the merging scale variation they also exhibit the largest nLO corrections (see K-factors in Fig.~\ref{Fig.: WZ nLO+PS validation continued}) compared to the other polarisation components. Thus, emissions between 20 and 40~GeV are at least partly responsible for these increased K-factors. The shape of the increased TL and LT K-factors for $\Delta \phi_{\text{e}^+\mu^-}$ in the COM polarisation definition as well as the giant TL and LT K-factors for large $p_{\perp, \mu^-}$ observed for both investigated polarisation definitions, however, can not be explained by contributions from emissions below 40~GeV. Here, hard emission corrections, beyond scales of 40\,GeV, contained in both nLO+PS and this multijet merged calculation, seem to play the dominant role.

	\section{Conclusions}
\label{sec:conclusions}

In this paper we presented a new implementation enabling the MC event generator \Sherpa to simulate polarised cross sections for an arbitrary number of intermediate vector bosons. The simulation itself remains unpolarised, comprising all polarisation components. However, the individual polarised cross sections are made available as additional event weights such that polarised cross sections for all possible polarisation combinations as well as all interferences between different polarisation states, which are made available here for the first time, can be calculated in a single simulation run. All common reference systems are supported, and new frames of interest can easily be implemented.

We validated our implementation at fixed LO revealing excellent agreement with the literature despite using different approximations to define the on-shell intermediate vector bosons (we use the narrow-width approximation, whereas the double-pole approximation was used in the reference results).
Using the example of inclusive $\text{W}^+\text{Z}$ production, we have shown that
our framework is not only able to calculate polarised cross sections at LO accuracy, but can also incorporate NLO corrections by using the new polarisation framework together with \Sherpa's \MCatNLO matching. In this way, we have presented the first nLO+PS matched predictions for polarised cross sections, showing important improvements over their only LO accurate counterparts available up until now.

Further, we have presented the combination of our new polarisation framework with \Sherpa's multijet-merging calculations, making polarised multijet calculations available. We have shown that, similarly to the unpolarised case, most features of the nLO accurate polarisation fractions, apart from a global normalisation factor, can be reproduced.
The additional systematics typically associated with this approach are negligible for integrated polarisation fractions, but can start to play a role for some subleading polarisation components in typical observables.

In the future, we plan to extend our polarisation framework to full NLO accuracy in the vector boson production part by including the complete virtual contributions in the calculation of the amplitude tensor. With that, we will also be able to include approximate NLO EW corrections in the \EWvirt\ \cite{Kallweit2015} scheme in the simulation of polarised cross sections with \Sherpa. A simulation of polarised cross sections with full NLO accuracy throughout requires the inclusion of higher order contributions in the decay part of the processes, which is more challenging and will follow in a second step.

	\section*{Acknowledgments}
The authors would like to thank Giovanni Pelliccioli for helpful discussions about \Phantom and the definition of polarised cross sections in general as well as Stefan Höche for his help with the polarisation definition in \Sherpa's build-in matrix element generator \Comix.\\
M.S.\ is supported by the UK Science and Technology Facilities Council (STFC)
Consolidated Grant programme ST/T001011/1 and by the Royal Society through a University
Research Fellowship (URF\textbackslash{}R1\textbackslash{}180549,  URF\textbackslash{}R\textbackslash{}231031) and
Enhancement Awards (RGF\textbackslash{}EA\textbackslash{}181033,
CEC19\textbackslash{}100349, and RF\textbackslash{}ERE\textbackslash{}210397).
This work has received funding from the European Union's Horizon 2020 research and innovation programme as part of the
Marie Skłodowska-Curie Innovative Training Network MCnetITN3 (grant agreement no. 722104).

	\appendix
	\section{Simulation of polarised cross sections with \Sherpa}
\subsection{Input structure for calculating polarised cross sections with \Sherpa} \label{Appendix: Input structure}
In this section, the structure of the run card parts relevant for simulating polarised cross sections with \Sherpa is discussed. In order to allow for the calculation of polarised cross sections, \texttt{Hard\_Decays} need to be enabled. Gauge invariance is retained by setting the VB widths to zero. This leads to real couplings.\\
The simulation of polarised cross sections in \Sherpa itself is steered by an own block called \texttt{Pol\_Cross\_Section} within the \texttt{Hard\_Decays} scoped setting in the \Sherpa run card. The following settings are possible:
\begin{description}
	\item[\texttt{Enabled}:] \texttt{<true/false> }enables the calculation of polarised cross sections on top of the simulation of unstable intermediate particles.
	\item[\texttt{Spin\_Basis}] specifies the spin basis. Beside the helicity basis (\texttt{Helicity}, default), constant reference vectors $a^\mu$ are supported,
	e.g.\ \texttt{1.0, 0.0, 0.0, 0.0}, which define the spin axis as shown in Eq.~\ref{Eq.: Spin axis}. Per simulation, only one spin basis can be used.
	\item[\texttt{Reference\_System}] denotes the reference system used for polarisation definition. Currently, the laboratory system (\texttt{Lab}, default), the center-of-mass frame of all intermediate particles (\texttt{COM}), the parton-parton-frame (\texttt{PPFr}) and the rest frames defined by any combination of the initial or final state particles in the VB production process are supported. The first three systems are specified by the corresponding keywords, for the latter the particle numbers (according to \Sherpa's internal particle numbering) of the particles defining the rest frame need to be given, separated by white spaces, e.g.\ \texttt{2 3}. In one simulation run, more than one reference system can be considered by passing a list of desired reference systems here.
	\item[\texttt{Transverse\_Weights\_Mode}] allows to switch between the coherent (mode \texttt{1}, default) and the incoherent (mode \texttt{0}) definition of transverse polarisation states (see Appendix~\ref{Appendix: Provided polarisation weights} for details), also weights for both definitions can be calculated at the same time (mode \texttt{2}).
	\item[\texttt{Weight<$n$>}] can be used to specify weights which should be calculated additionally to the base polarisation weights which are output during each simulation run, see Appendix~\ref{Appendix: Provided polarisation weights} for details. \texttt{<$n$>} needs to be replaced by an integer. By using different integers, more than one custom weight can be calculated.
\end{description}
With that, the parts of the run card relevant for the simulation of polarised cross sections e.g. in Sec.~\ref{sec:pheno} are:
\par\noindent\rule{\textwidth}{0.4pt}
{
\begin{verbatim}
WIDTH_SCHEME: Fixed  # real couplings

# Gauge of the Weyl spinors to receive the representation of the polarisation vectors 
# in Eq. (2.4)
COMIX_DEFAULT_GAUGE: 0

PARTICLE_DATA:
  24: {Width: 0}
  23: {Width: 0}

HARD_DECAYS:
  Enabled: true
  Pol_Cross_Section:
    Enabled: true
    Spin_Basis: Helicity    # default
    Reference_System: [Lab, COM]
    Transverse_Weights_Mode: 1  # default
    Weight1: 2
    Weight2: 3

\end{verbatim}}
\par\noindent\rule{\textwidth}{0.4pt}
    
\subsection{Output structure: provided polarisation weights} \label{Appendix: Provided polarisation weights}
As introduced in Sec.~\ref{Sec.: Polarised cross sections with Sherpa}, polarised cross sections of all possible polarisation combinations of the intermediate particles are provided as additional event weights in \Sherpa. There are three different categories of polarisation weights in \Sherpa:
\begin{description}
	\item[Base polarisation weights] are output during each simulation of polarised cross sections and include all contributions where all intermediate VBs are in a defined polarisation state (for VBs: left(-)-handed, right(+)-handed or longitudinal (0) polarisation mode). Corresponding weight names have the form \\ \texttt{PolWeight\_<Referencesystem>.particle1.$\lambda_1$\_particle2.$\lambda_2$$\ldots$} with $\lambda_i \in \{+,-,0\}$. The order of the particles in the weight names is determined by \Sherpa's internal particle ordering. Furthermore, the all interference terms are totalled to an overall interference contribution with the label \\ \texttt{PolWeight\_<Referencesystem>.int}. 
	\item[Transversely polarised weights] describing contributions of polarisation combinations where at least one massive VB is transversely polarised are also output per default. Two distinct definitions for the transverse polarisation are supported:
	\begin{itemize}
		\item \textbf{incoherent definition:} left- and right-handed polarised contributions are added to a transverse contribution. Corresponding weights contain a small \textbf{``t''} for each transverse polarised particle, e.g. \texttt{PolWeight\_Lab.W+.t}. 
		\item \textbf{coherent definition:} besides the left- and right-handed polarised contributions also left-right-interference terms are included in the definition of the transverse contribution. This definition is more common in the literature \cite{Ballestrero2017, Ballestrero2019, Ballestrero2020, Denner2020, Denner2021, Denner2021a} and is also used by other generators, e.g. \Madgraph. Thus, it is also the default choice in \Sherpa. If this definition is chosen, also an adjusted interference weight is calculated with the weight name \texttt{PolWeight\_<Referencesystem>.coint}. Corresponding weights are recognizable by a capital \textbf{``T''} for each transverse polarised particle, e.g. \texttt{PolWeight\_Lab.W+.T}.
	\end{itemize} 
	\item[Custom polarisation weights] are only calculated if specified in the \Sherpa run card. The corresponding setting is \texttt{Weight<$n$>} where $n$ is an integer such that it is possible to request more than one custom weight. Depending on the type of custom weights either weight names or particle numbers used to specify which weight should be calculated additionally. The following custom weights can be provided by \Sherpa:
	\begin{itemize}[leftmargin=0cm]
		\item \textbf{partially unpolarised weights:} Intermediate particles, that shall be considered as unpolarised, can be specified by a comma-separated list of their particle numbers according to \Sherpa's internal particle numbering in the run card. The associated weight names have the form \\ \texttt{PolWeight\_<Referencesystem>.Weight<$n$>\_particle1.U\_particle2.U...\_particlei.$\lambda_i$...} \\ where particles $1-(i-1)$ are considered as unpolarised. \texttt{Weight<$n$>} corresponds to the setting name in the \Sherpa run card to distinguish between different sets of unpolarised particles. The ordering of the unpolarised and the polarised particles among itself is again according to \Sherpa's internal particle ordering. In addition, a new interference weight is calculated \\ (\texttt{PolWeight\_<Referencesystem>.Weight<$n$>\_particle1.U\_particle2.U...\_int}). It contains less terms than the interference corresponding to the base weights, since the polarisation weights in which the remaining polarised intermediate particles have a definite polarisation also contain interference terms from the now unpolarised particle. For example, if a process with two VBs is considered and the VB with the polarisation indices $\lambda_1$, $\lambda_1^\prime$ in Eq.~\ref{Eq.: Polarised matrix element} remains unpolarised then the following entries of the amplitude tensor would contribute to the resulting single-polarised (=one of two intermediate VBs is in a definite polarisation state) longitudinal weight with the interference terms in square brackets:\\
		$|\mathcal{M_{\text{singlepol}}}|^2_{00} =|\mathcal{M}|^2_{+0+0}+|\mathcal{M}|^2_{-0-0}+|\mathcal{M}|^2_{0000} + \Big[ |\mathcal{M}|^2_{+0-0}+|\mathcal{M}|^2_{+000}+|\mathcal{M}|^2_{-0+0}+|\mathcal{M}|^2_{-000}+|\mathcal{M}|^2_{00+0}+|\mathcal{M}|^2_{00-0} \Big]$.
		\item \textbf{individual interference weights:} Interference weight names have two instead of one polarisation index per particle (first index stands for the polarisation of the particle in the corresponding matrix element, the second index for its polarisation in the complex conjugate matrix element). If provided within the \texttt{Weight<$n$>} setting, the corresponding interference weight is output and labelled by its weight name. 	
		\item \textbf{sum of specified weights:} All weights from those mentioned beforehand, which are specified as comma separated list of their weight names in the run card, are totalled and labelled as \\ \texttt{PolWeight\_<Referencesystem>.Weight<$n$>} where \texttt{Weight<$n$>} is again the corresponding setting in the \Sherpa run card. 
	\end{itemize}
\end{description}

\section{Validation setups and further results}
\label{Appendix: validation simulation setup}

This appendix summarises the simulation setups used for the validation study presented in Sec.~\ref{Sec.: Validation} in Sec.~\ref{Appendix:Simulation setup validation}. Furthermore, some additional validation results are given in Sec.~\ref{Appendix.: Additional validation results}.

\subsection{Validation setups}
\label{Appendix:Simulation setup validation}

In order to reproduce the results from the literature studies in Ref.s~\cite{Ballestrero2017, Ballestrero2019, Ballestrero2020}, the simulation parameters and phase space definitions given in Tabs.~\ref{Tab.: simulation settings validation study} and \ref{Tab.: selection criteria for validation study}, respectively are chosen identical to the literature. Polarised cross sections with and without applying lepton acceptance criteria are investigated and denoted as \textbf{``fiducial''} and \textbf{``inclusive''} setup, respectively. All fermions except the top quark are considered as massless, if not stated otherwise. A b-veto is understood as a perfect b-veto on initial and final states. The cross sections studied, whether single- or double-polarised, and the definitions of polarisation investigated, match those found in the literature.  
\begin{table}[t!]	
	\renewcommand{\arraystretch}{1.3}
	\centering
	\small
	\begin{tabular}{ll}
		\toprule
		PDF-Set from LHAPDF6~\cite{Buckley2015}  & NNPDF30\_lo\_as\_0130~\cite{NNPDFCollaboration2014} \\
		Electroweak scheme &  $G_\mu$ scheme with $G_\mu=1.16637\cdot 10^{-5}$~$\text{GeV}^{-2}$ and
		$\alpha_{\text{EW}}=\frac{\sqrt{2}G_\mu M^2_\text{W}}{\pi}\Big(1-\Big(\frac{M_\text{W}}{M_\text{Z}}\Big)^2\Big)$ \\
		& complex mass scheme (full), real EW parameters (polarisation) \cite{Ballestrero2017}\\
		Strong coupling $\alpha_S(M^2_\text{Z})$ & 0.130\\
		Factorization scale & \cite{Ballestrero2017, Ballestrero2019}: $\mu_F = M_{4\text{l}}/\sqrt{2}$, \cite{Ballestrero2020}: $\mu_F = \sqrt{p_{\perp,\text{j}_1}p_{\perp, \text{j}_2}}$\\
		VB pol masses &$M_\text{W}=80.358 \text{ GeV}$, $M_\text{Z}=91.153 \text{ GeV}$\\
		VB pol widths & zero for all VBs in polarisation calculations, \\
		&$\Gamma_\text{W}=2.084 \text{ GeV}$,  $\Gamma_\text{Z}=2.494 \text{ GeV}$ otherwise\\
		\bottomrule
	\end{tabular}
	\caption{Simulation settings used for \Sherpa simulations within the validation study of the new \Sherpa polarisation framework against literature data.}
	\label{Tab.: simulation settings validation study}
\end{table}
\begin{table}[t!]
	\centering
	\renewcommand{\arraystretch}{1.3}
	\small
	\begin{tabular}{ll}
		\toprule
		\multicolumn{2}{c}{General selection requirements}\\
		inclusive phase space & fiducial phase space \\
		\midrule
		$|\eta_{\text{j}}|<5$ & cf. inclusive phase space \\
		$p_{\perp, \text{j}} > 20$ GeV & $|\eta_\text{l}| < 2.5$\\
		$M_{\text{jj}} > 500$ GeV & $p_{\perp, \text{l}} > 20$ GeV \\
		$|\Delta\eta_{\text{jj}}| > 2.5$ & $\text{W}^+$Vjj: $p_{\perp, \text{miss}}>40$ GeV \\
		\hline
		\multicolumn{2}{c}{Process specific selection requirements}\\
		process (reference) & selection criteria \\
		\hline
		$\mathrm{W}^+$$\mathrm{W}^+$jj \cite{Ballestrero2020}: pp $\rightarrow$ $\mathrm{e^+}$ $\nu_e$ $\mu^+\nu_\mu$jj & $M_{4l}>161$ GeV \tablefootnote{The $M_{4l}$ selection requirement for the $\mathrm{W}^+$$\mathrm{W}^+$jj process is only applied during the polarisation calculation.} \\
		\hline
		$\mathrm{W}^+$$\mathrm{W}^-$jj \cite{Ballestrero2020}: pp $\rightarrow$ $\mathrm{e^+}$ $\nu_\text{e}$ $\mu^-$ $\bar{\nu}_\mu$jj & b-veto, $M_{4\text{l}}>2M_\text{W}$ \\
		$\mathrm{W}^+$$\mathrm{W}^-$jj \cite{Ballestrero2017}: pp $\rightarrow$ $\mathrm{\mu^+}$ $\nu_\mu$ $\text{e}^-$ $\bar{\nu}_\text{e}$jj & b-veto, $M_{\text{4l}}> 300$ GeV, lepton selection criteria only on $\text{e}^-$ \\
		&$M_{\text{jj}}>600$ GeV, $|\Delta \eta_{\text{jj}}|>3.6$, $\eta_{\text{j}_1}\cdot\eta_{\text{j}_2}$<0\\
		\hline
		$\mathrm{W}^{+}$Zjj \cite{Ballestrero2020}: pp $\rightarrow$ $\mathrm{e^+}$ $\nu_\text{e}$ $\mu^+\mu^-$jj & b-veto, $M_{4\text{l}}>200$ GeV, $|M_{\text{e}^+\text{e}^-}-M_\text{Z}|< 10$ GeV\\
		$\mathrm{W}^{+}$Zjj \cite{Ballestrero2019}: pp $\rightarrow \mu^+\nu_\mu \mathrm{e}^+\text{e}^- $jj & b-veto, $M_{\text{4l}}>200$ GeV, $|M_{\text{e}^+\text{e}^-}-M_\text{Z}|< 15$ GeV\\
		\hline
		ZZjj \cite{Ballestrero2020}: pp $\rightarrow$ $\mathrm{e^+}$ $\mathrm{e}^-$ $\mu^+\mu^-$jj & b-veto, $M_{4\text{l}}>200$ GeV, $|M_{\text{l}^+\text{l}^-}-M_\text{Z}|< 10$ GeV\\
		ZZjj \cite{Ballestrero2019}: pp $\rightarrow$ $\mathrm{e^+}$ $\mathrm{e}^-$ $\mu^+\mu^-$jj & b-veto, $M_{4\text{l}} > 200$ GeV, $|M_{\text{l}^+\text{l}^-}-M_\text{Z}|< 15$ GeV \\
		\bottomrule
	\end{tabular}
	\caption{Selection criteria used for \Sherpa simulations within the validation study of the new \Sherpa polarisation framework against literature data.}
	\label{Tab.: selection criteria for validation study}
\end{table}   

\Rivet analyses used to analyse the simulation data are based on the ATLAS analysis of Ref.~\cite{ATLASCollaboration2017} (ZZjj), on the \Rivet analyses of Ref.~\cite{Bittrich2022} ($\text{W}^+\text{W}^+$jj) and Ref.~\cite{Burghardt2022} ($\text{W}^+$Zjj validation against Ref.~\cite{Ballestrero2019}) or on private \Rivet analyses ($\text{W}^+$Zjj validation against Ref.~\cite{Ballestrero2020}, $\text{W}^+\text{W}^-$jj).
The study of the $\text{W}^+$Zjj process in Ref.~\cite{Ballestrero2019} uses a dedicated procedure to reconstruct neutrinos for the calculation of $\text{W}^\pm$~boson observables which is taken into account in the associated \Rivet analysis. For all other process, perfect neutrino reconstruction is assumed.
Within the event generation, only selection criteria on final state particles of the hard scattering process (= VB production subprocess) can be applied. All remaining selection criteria are then implemented in the subsequent \Rivet-analysis.

For all processes, observables are investigated that can only be defined for the mixed lepton flavour decay channel of the VB pair if the VBs are identical. Same flavour lepton decay channels which can not be excluded by \Sherpa during the simulation of several identical VBs (ZZjj, $\text{W}^+$$\text{W}^+$jj process), are then vetoed during the \Rivet-analyses for all observables. Therefore, no ambiguity for the reconstruction of the VBs from their decay products exist. Corresponding single-polarised cross sections are calculated by a private extension of the presented polarisation framework, where all polarisation combinations are totalled, leaving only the VB decaying via a specified decay channel polarised. The resulting weights are set to zero for same flavour decay channels.

\subsection{Integrated polarised cross sections in the presence of lepton acceptance requirements} \label{Appendix.: Additional validation results}
This appendix presents integrated cross sections computed in the fiducial phase spaces defined in Sec.~\ref{Appendix:Simulation setup validation} to supplement the distributions discussed in Sec.~\ref{Sec.: Validation}. Resulting integrated cross sections are displayed in Tab.s~\ref{Tab.:ZZjj-validation-total cross section fiducial phase space}-\ref{Tab.: ssWW validation fiducial phase space}. Interference predictions from the literature are calculated as difference between the unpolarised result (unpol) and the sum of all polarised contributions (polsum) for comparison. 

For all four processes, the unpolarised result reproduces the full cross section at the 1\% level, indicating small non-resonant contributions and low off-shell effects not covered by the mass smearing.
\begin{table}[t!]
	\renewcommand{\arraystretch}{1.3}
	\centering
	\small
	\begin{tabular}{lllll}
		\toprule
		ZZjj& $\sigma_{\text{\Phantom}}$  [fb] & Fraction [\%] &  $\sigma_{\text{\Sherpa}}$ [fb] & Fraction [\%]\\
		\midrule
		full & 0.06102(4) && 0.060987(27) &\\
		unpol & 0.06059(4) & 100 & 0.06116(4) & 100\\
		\hline
		polsum & 0.05891(2) & 97.23(7) & 0.059452(29) & 97.21(8)\\
		int & 0.00168(4) (diff) & 2.77(7) & 0.001712(24) & 2.80(4)\\
		\hline
		long-unpol & 0.01619(1) & 26.721(24) & 0.016230(15) & 26.536(29) \\
		left(-)-unpol & 0.02676(2) & 44.17(4) & 0.027079(21) & 44.27(4)\\
		right(+)-unpol & 0.01595(1) & 26.324(24) & 0.016143(14)  & 26.394(28)\\
		\bottomrule
	\end{tabular}
	\caption{Integrated single-polarised cross sections for the ZZjj process at fixed LO obtained with \Phantom in Ref.~\cite{Ballestrero2019} and \Sherpa in the fiducial phase space introduced in Sec.~\ref{Appendix:Simulation setup validation}; single-polarised cross sections are given for the Z boson decaying into an electron-positron pair being polarised; the polarisation is defined in the laboratory frame; polarisation fractions are calculated relative to the unpolarised result.}
	\label{Tab.:ZZjj-validation-total cross section fiducial phase space}
\end{table}
\begin{table}[t!]
	\renewcommand{\arraystretch}{1.3}
	\centering
	\small
	\begin{tabular}{lllll}
		\toprule
		$\text{W}^+\text{W}^-$jj& $\sigma_{\text{\Phantom}}$  [fb] & Fraction [\%] &  $\sigma_{\text{\Sherpa}}$ [fb] & Fraction [\%]\\	
		\midrule
		full & 1.411(1) & 100 & 1.3537(11) & 100 \\
		unpol & 1.401(1)& 99.29(10) & 1.3497(5) & 99.70(9)\\
		\hline
		polsum & 1.382(1) & 97.94(10) & 1.3309(9)& 98.32(11)\\
		int & 0.019(1) (diff) & 1.35(7) & 0.0188(8)& 1.39(6)\\
		\hline
		long & & 21& 0.28969(33)& 21.400(30)\\
		left(-) & & 52 & 0.7047(5)& 52.06(6)\\
		right(+) && 25 & 0.3365(4)& 24.86(4)\\
		\bottomrule
	\end{tabular}
	\caption{Integrated cross sections for the $\text{W}^+\text{W}^-$jj process at fixed LO obtained with \Phantom in Ref.~\cite{Ballestrero2017} and \Sherpa in the fiducial phase space introduced in Sec.~\ref{Appendix:Simulation setup validation}; single-polarised cross sections are given for the $\text{W}^-$ boson being polarised; the polarisation is defined in the laboratory frame; polarisation fractions are calculated relative to the full result.}
	\label{Tab.: W+W- integrated cross sections, fiducial phase space}
\end{table}
For the $\text{W}^+\text{W}^-$jj process, it was not possible to achieve a perfect agreement of the full results from \Sherpa and \Phantom. However, since the literature only provides polarisation fractions (relative to the full!) for this process, this does not pose a limitation for the comparison. 

If given, polarised cross sections, polarisation and interference fractions obtained with \Sherpa are in very good agreement with the literature for all processes showing deviations of 1.5\% or less. 
\begin{table}[t!]
	\renewcommand{\arraystretch}{1.3}
	\small
	\centering
	\begin{tabular}{lllll}
		\toprule
		$\text{W}^+\text{Z}$jj  & $\sigma_{\text{\Phantom}}$ [fb]   & Fraction [\%] & $\sigma_{\text{\Sherpa}}$ [fb]   & Fraction [\%]   \\
		\midrule
		full            & 0.1651(1) &       & 0.16519(5)                             &             \\
		unpol                  & 0.1642(2) & 100 & 0.16342(9)                & 100            \\
		\hline
		\multicolumn{5}{c}{Z boson polarised}\\
		\hline
		int                    & 0.0040(2) (diff) & 2.43(12) & 0.00397(7)                & 2.43(4)     \\
		right(+)               & 0.04054(3) & 24.689(35) & 0.04042(4)                & 24.734(28)  \\
		left(-)                & 0.07687(6) & 46.81(7) & 0.07657(5)                & 46.86(4)    \\
		long                   & 0.04256(3) & 25.920(36) & 0.04246(4)                & 25.980(27)  \\
		\hline
		\multicolumn{5}{c}{$\text{W}^+$ boson polarised}\\
		\hline
		int                    & 0.0038(2) (diff) & 2.31(12) & 0.00362(7)                & 2.22(5)     \\
		right(+)               & 0.03093(2) & 18.837(26) & 0.03085(4)                & 18.880(26)  \\
		left(-)                & 0.09631(8) & 58.65(9) & 0.09602(7)                & 58.75(5)    \\
		long                   & 0.03321(3) & 20.225(31) & 0.03293(4)                & 20.148(25)  \\
		\bottomrule
	\end{tabular}
	\caption{Integrated single- and double-polarised cross sections for the $\text{W}^+$$\text{W}^+$jj process at fixed LO obtained with \Phantom in Ref.~\cite{Ballestrero2020} and \Sherpa in the fiducial phase space introduced in Sec.~\ref{Appendix:Simulation setup validation}; single-polarised cross sections are given for the $\text{W}^+$ boson decaying into $\text{e}^+\nu_{\text{e}}$ being polarised; the polarisation is defined in the laboratory and the $\text{W}^+$$\text{W}^+$-center-of-mass frame; polarisation fractions are calculated relative to the unpolarised result.}
	\label{Tab.: WZjj validation fiducial phase space}
\end{table}
\begin{table}[t!]
	\renewcommand{\arraystretch}{1.3}
	\small
	\centering
	\begin{tabular}{lllll}
		\toprule
		$\text{W}^+\text{W}^+$jj   & $\sigma_{\text{\Phantom}}$ [fb]   & Fraction [\%]   & $\sigma_{\text{\Sherpa}}$ [fb]   & Fraction [\%]   \\
		\midrule
		full                     & 1.593(2)                       &             &      1.5901(13)                           &             \\
		unpol                   & 1.572(2)                       &    100    & 1.5728(8)   &   100    \\
		\hline
		\multicolumn{5}{c}{Laboratory frame}\\
		\hline
		int (single)                      & -0.0156(23) (diff)                    & -0.99(15)   & -0.0141(4)                      & -0.898(26)  \\
		T-unpol                  & 1.165(1)                       & 74.11(11)   & 1.1646(6)                       & 74.05(5)    \\
		L-unpol                  & 0.4226(4)                      & 26.88(4)    & 0.4223(4)                       & 26.851(27)  \\
		\hline
		int (double)                     & -0.0279(22) (diff)                   & -1.77(14)   & -0.0281(7)                      & -1.78(4)    \\
		L-L                      & 0.1185(1)                      & 7.538(12)   & 0.11837(19)                     & 7.526(13)   \\
		TL+LT                    & 0.6124(6)                      & 38.96(6)    & 0.6134(5)                       & 39.00(4)    \\
		T-T                      & 0.8690(9)                      & 55.28(9)    & 0.8691(6)                       & 55.26(5)    \\
		\hline
		\multicolumn{5}{c}{$\text{W}^+\text{W}^+$-center-of-mass frame}\\
		\hline
		int (single)                     & -0.0136(29) (diff)           & -0.87(18)   & -0.0118(4)                      & -0.749(26)  \\
		T-unpol                  & 1.182(2)                       & 75.19(16)   & 1.1807(7)                       & 75.07(6)    \\
		L-unpol                  & 0.4036(5)                      & 25.67(5)    & 0.40393(34)                     & 25.682(25)  \\
		\hline
		int (double)                      & -0.0220(22) (diff)                 & -1.40(14)   & -0.0213(6)                      & -1.35(4)    \\
		L-L                      & 0.1552(2)                      & 9.873(18)   & 0.15593(21)                     & 9.914(14)   \\
		TL+LT                    & 0.5038(6)                      & 32.05(6)    & 0.5046(4)                       & 32.084(31)  \\
		T-T                      & 0.9350(9)                      & 59.48(9)    & 0.9338(6)                       & 59.37(5)    \\
		\bottomrule
	\end{tabular}
	\caption{Integrated single-polarised cross sections for the $\text{W}^+$$\text{Z}$jj process at fixed LO obtained with \Phantom in Ref.~\cite{Ballestrero2019} and \Sherpa in the fiducial phase space introduced in Sec.~\ref{Appendix:Simulation setup validation}; the polarisation is defined in the laboratory frame; polarisation fractions are calculated relative to the unpolarised result.}
	\label{Tab.: ssWW validation fiducial phase space}
\end{table}

	\printbibliography	
\end{document}